\documentclass[a4paper,11pt]{article}
\pdfoutput=1 
\usepackage{jcappub} 
\usepackage{color,soul}
\usepackage{float}
\usepackage{graphicx}
\usepackage{bm}
\usepackage{epsfig}
\usepackage{subfigure}
\usepackage{latexsym, amssymb}
\usepackage[colorlinks=true,linkcolor=blue,citecolor=blue]{hyperref}

\title{Primordial Black Holes Dark Matter and Secondary Gravitational Waves from Warm Higgs-G Inflation} 	
\author[a]{Richa Arya,}
\emailAdd{richaarya@iisc.ac.in} 
\affiliation[a]{Department of Physics, 
	Indian Institute of Science, C. V. Raman Road, Bangalore 560012, India}
\author[a]{Rajeev Kumar Jain,}
\emailAdd{rkjain@iisc.ac.in} 
\author[b]{Arvind Kumar Mishra} 
\emailAdd{arvind.mishra@acads.iiserpune.ac.in}
\affiliation[b]{Indian Institute of Science Education and Research, Pune 411008, India}

\date{\today}

\abstract{
	We explore the role of dissipative effects during warm inflation leading to the small-scale enhancement of the power spectrum of curvature perturbations. In this paper, we specifically focus on non-canonical warm inflationary scenarios and study a model of warm Higgs-G inflation, in which the Standard Model Higgs boson drives inflation, with a Galileon-like non-linear kinetic term. 
	We show that in the Galileon-dominated regime, the primordial power spectrum is strongly enhanced, leading to the formation of primordial black holes (PBH) with a wide range of the mass spectrum.
	Interestingly,  PBHs in the asteroid mass window $\sim (10^{17}$ -- $10^{23}$) g are generated in this model, which can explain the total abundance of dark matter in the Universe.
	In our analysis, we also calculate the secondary gravitational waves (GW) sourced by these small-scale overdense fluctuations and find that the induced GW spectrum can be detected in future GW detectors, such as LISA, BBO, DECIGO, etc. Our scenario thus provides a novel way of generating PBHs as dark matter and a detectable stochastic GW background from warm inflation. We also show that our scenario is consistent with the swampland and the trans-Planckian censorship conjectures and, thus, remains in the viable landscape of UV complete theories.}

\begin{document}
	\maketitle
	\flushbottom

	\section{Introduction}
	The inflationary paradigm \cite{Kazanas:1980tx,Sato:1980yn,Guth:1980zm,Linde:1981mu} of the early Universe successfully explains the  observations of anisotropies in the cosmic microwave background (CMB) radiation and generates seed inhomogeneties for the large scale structure (LSS) formation \cite{Planck:2018jri} (for reviews, see \cite{Baumann:2009ds,Linde:2007fr,Riotto:2002yw,Tsujikawa:2003jp}).
	Despite its phenomenal success, the underlying particle physics model of inflation remains elusive to date \cite{Linde:2005ht}. In the Standard Model (SM) of particle physics, the only scalar field is the Higgs boson, whose self-interactions are so strong that it can not act as an inflaton (a scalar field that drives inflation) in a minimal setup and thus, can not explain the cosmological observations. However, by extending it to a non-minimal configuration, 
	one can construct a viable inflationary model within the SM (for a review, see Ref. \cite{Rubio:2018ogq}). This can be achieved, for example, by introducing a non-minimal coupling of the Higgs field to gravity \cite{Bezrukov:2007ep,Barvinsky:2008ia,Germani:2010gm}, or with a non-canonical kinetic term of the Higgs field, such as in the $k$-inflation \cite{Armendariz-Picon:1999hyi}, ghost condensate \cite{Arkani-Hamed:2003juy}, and Dirac-Born-Infeld inflation \cite{Alishahiha:2004eh} models. 
	
	One such construction with a general kinetic term and additional non-linear terms are the Galileon-like models whose Lagrangian can be written as
	\begin{equation}
		\mathcal{L}_\phi=K(\phi,X)-G(\phi,X)\Box \phi,
		\label{nonmin}
	\end{equation}
	where $X=-\frac{1}{2} g^{\mu\nu} \partial_\mu\phi \partial_\nu\phi$ is the standard kinetic term of the Galileon field $\phi$, $K(\phi,X)=X-V(\phi)$ where $V(\phi)$ is the field potential, and $G(\phi,X)$ is an arbitrary function of $\phi, X$. This specific form of the non-linear kinetic term is special, as it does not lead to extra degrees of freedom or ghost instabilities and maintains the gravitational and scalar field equations at the second order \cite{Deffayet:2009wt,Deffayet:2009mn,Kobayashi:2011nu}. The Galileon field is named so because it possesses a Galileon shift symmetry in the Minkowski spacetime. Phenomenologically, Galileon-type scalar fields have been extensively studied in the context of dark energy, modified gravity \cite{Chow:2009fm,Silva:2009km,Kobayashi:2009wr,Kobayashi:2010wa,Gannouji:2010au,DeFelice:2010gb,DeFelice:2010pv} as well as inflation \cite{Kobayashi:2010cm,Mizuno:2010ag,Burrage:2010cu,Creminelli:2010qf,Ohashi:2012wf,Choudhury:2012whm}.  In Ref. \cite{Kamada:2010qe}, the authors explored Higgs inflation in the presence of Galileon-like non-linear derivative interactions and demonstrated that this model is compatible with cosmological observations. Also, the tensor-to-scalar ratio for this model was found to be within the sensitivity of future experiments, which can be used to discriminate it from the standard inflationary model with a canonical kinetic term. However, when the Galileon term dominates over the standard kinetic term in this model, the dynamics of reheating is modified. For large self-coupling $\lambda$, there is no oscillatory phase and moreover, the square of sound speed $c_s^2$ is negative during reheating, leading to instabilities of small-scale perturbations \cite{Ohashi:2012wf}. 
	In order to alleviate this problem, the authors in Ref. \cite{Kamada:2013bia} extended the Higgs Lagrangian with higher order kinetic terms to obtain $c_s^2 >0$, thereby avoiding instabilities. 
	Alternatively, this problem would not arise in the first place if there is no separate reheating phase after inflation, as it could be in the case of warm inflation. This was explored by the authors in Refs. \cite{Motaharfar:2017dxh,Herrera:2017qux,Motaharfar:2018mni}, and forms the basis of this paper.

	Warm inflation \cite{Berera:1995wh,Berera:1995ie,Berera:1998px} is a general description of inflation in which the dissipative effects during inflation play an important role in the inflationary dynamics.
	The basic idea of warm inflation is that the inflaton is sufficiently coupled with other fields, such that it dissipates its energy into them during its evolution, which leads to a thermal bath (with temperature $T$) in the Universe during the inflationary phase.
	Therefore, a separate reheating phase is not necessarily needed for particle production. For reviews on warm inflation, see Refs. \cite{Berera:2006xq,Berera:2008ar,Oyvind&nbspGron:2016zhz}. 
	The background dynamics of the inflaton as well as its fluctuations are modified in warm inflation, therefore, the primordial power spectrum has distinct signatures on cosmological observables, as compared to cold inflation. For instance, the tensor-to-scalar ratio in warm inflation is lowered, thus certain models of cold inflation, although ruled out from Planck observations, could be viable models in the warm inflation description \cite{Visinelli:2016rhn,Benetti:2016jhf,Arya:2017zlb,Arya:2018sgw,Bastero-Gil:2017wwl}. Warm inflationary models also predict unique non-Gaussian signatures that can be used to test these models \cite{Gupta:2002kn,Moss:2011qc,Bastero-Gil:2014raa, Mirbabayi:2022cbt}. As the inflationary dynamics is modified in warm inflation, the slow roll conditions which demand an extremely flat potential, are also relaxed in these models \cite{Berera:2004vm}. Further, some warm inflation models can provide a unified description for inflation, dark matter, and/or dark energy \cite{Rosa:2018iff,Rosa:2019jci,Sa:2020fvn,DAgostino:2021vvv}. Another interesting aspect of warm inflation models is that they can also explain the baryon asymmetry of the Universe \cite{Bastero-Gil:2011clw,Basak:2021cgk}. Warm inflation studies also show that for
	some models with a large value of the dissipation parameter, the swampland  conjectures can be satisfied, thus making them in agreement with a high energy theory \cite{Das:2018rpg,Motaharfar:2018zyb}.  We will discuss this aspect of our model in detail in Section \ref{WHGI}. Since all these  features of warm inflation arise from the fundamental principles of a dissipating system, it is very crucial to study warm inflation to understand the physics of the early Universe.
	
	Although the large scale imprints of warm inflation are well studied through the  observations of the CMB anisotropies, the small scale features have recently acquired much attention. One of the novel probes of small scale physics of inflation is the formation and abundance of primordial black holes (PBH) \cite{zeldo:1966,Hawking:1971ei,Carr:1974nx}. In contrast to the astrophysical black holes, which are the end stages of a star, PBHs are primordial in origin and may form by different mechanisms, such as the collapse of overdense fluctuations \cite{Carr:1974nx,Carr:1975qj}, bubble collisions \cite{Hawking:1982ga}, collapse of strings \cite{Hogan:1984zb}, domain walls \cite{Caldwell:1996pt}, etc.  For reviews on PBH, see Refs. \cite{Khlopov:2008qy,Escriva:2022duf,Carr:2020xqk,Green:2020jor}. In order to produce PBHs by the gravitational collapse, the amplitude of primordial curvature power spectrum at small scales has to be 
	$\sim\mathcal{O}(10^{-2})$. Such an enhancement by several orders can be achieved in different  models, such as the hybrid inflation
	\cite{GarciaBellido:1996qt,Bugaev:2011wy,Clesse:2015wea}, running-mass inflation  \cite{Leach:2000ea,Kohri:2007qn,Drees:2011hb,Motohashi:2017kbs}, hilltop inflation \cite{Alabidi:2009bk}, inflating curvaton  \cite{Kohri:2012yw}, axion curvaton inflation \cite{Kawasaki:2012wr,Ando:2017veq}, double inflation \cite{Kawasaki:1997ju}, 
	thermal inflation \cite{Dimopoulos:2019wew}, multifield inflation \cite{Pi:2017gih,Braglia:2020eai}, single field inflation with a broken scale invariance \cite{Bringmann:2001yp}, or by introducing an inflection point in the potential 
	\cite{Garcia-Bellido:2017mdw,Ballesteros:2017fsr,Germani:2017bcs,Bhaumik:2019tvl,Ragavendra:2020sop,Karam:2022nym,Gu:2022pbo,Ragavendra:2023ret}, or bump/dip in the potential \cite{Mishra:2019pzq}, running of the spectral index \cite{Drees:2011yz, Kohri:2018qtx,Ghosh:2022okj}, suppression of the sound speed \cite{Ozsoy:2018flq,Zhai:2022mpi}, modified dispersion relations \cite{Ashoorioon:2019xqc}, or resonant instability \cite{Cai:2018tuh,Cai:2019bmk,Zhou:2020kkf}, modified gravity theories \cite{Kawai:2021edk,Lin:2021vwc,Papanikolaou:2021uhe,Papanikolaou:2022hkg}, etc.
	PBHs are crucial probes of the small scale features of the primordial power spectrum and hence different inflationary models \cite{Green:1997sz,Josan:2009qn,Carr:2009jm,Sato-Polito:2019hws}. 
	
	The abundance of PBHs is constrained through various observations, e.g. PBHs with mass $M_{PBH}<10^{15}$ g, would have evaporated by today, and thus their consequences on
	the Big-Bang nucleosynthesis (BBN) can provide constraints on their initial mass fraction. The PBHs with mass $M_{PBH} \lesssim 10^9$ g would completely evaporate by BBN, and therefore the bounds on their abundance are not very stringent. Recent studies show that such ultralight PBHs might induce interesting observational imprints such as an extra contribution to the dark radiation and dark relics \cite{Bhaumik:2022zdd}. 
	It is also possible that these PBHs may dominate the universe for a short duration before the radiation dominated epoch, and lead to a secondary resonant enhancement of the induced GWs \cite{Bhaumik:2020dor, Bhaumik:2022pil,Gehrman:2022imk}. 
	For PBHs with $M_{PBH}>10^{15}$ g, the constraints arise from their gravitational effects, like lensing, dynamical effects on interaction with astrophysical systems,   LIGO/Virgo gravitational wave (GW) merger events, etc. (For details, see Refs. \cite{Sasaki:2018dmp,Carr:2020gox}).  Also, PBHs can constitute a significant or total fraction of the dark matter (DM) density in the mass window ($10^{17}-10^{23}$) g (see recent reviews \cite{Carr:2020xqk,Green:2020jor}). Thus, PBHs are very important from the aspects of DM phenomenology. Further, as for PBH generation, the amplitude of scalar fluctuations at smaller scales is hugely enhanced, there is an inevitable GW spectrum sourced by these large density fluctuations \cite{Mollerach:2003nq,Ananda:2006af,Baumann:2007zm,Espinosa:2018eve,Kohri:2018awv}, which can have interesting observational consequences in the future GW detectors.
	
	Motivated by this, we investigated the small scale features of warm inflation in our earlier works \cite{Arya:2019wck,Arya:2022xzc} and considered minimally coupled single-field models with a canonical kinetic term. We explored the formation of PBHs and the scalar induced GW spectrum from a model of warm inflation in Refs. \cite{Arya:2019wck,Arya:2022xzc}. In our analysis, we found that for our model, the primordial power spectrum is red-tilted ($n_s<1$) for the CMB scales, but turns blue-tilted ($n_s>1$) at the small scales with a large amplitude. This generates PBHs of mass nearly $10^3$ g and an associated GW spectrum over the frequency range $(1-10^6)$ Hz. Further, these tiny mass PBHs would have evaporated by today, but the calculated initial abundance of these PBHs favours the possibility that the Planck-mass remnants of these PBHs could constitute the total dark matter.

	A similar analysis was performed in another theoretically interesting scalar warm little inflaton model \cite{Bastero-Gil:2021fac}.
	In this model, the dissipation coefficient $\Upsilon$, which is a measure of the inflaton interactions with other fields, switches its behaviour from $\Upsilon\propto 1/T$ to $\Upsilon\propto T^\kappa$ ($\kappa>0$) as inflation proceeds. It is found that the very small scale modes grow sufficiently to generate PBHs of mass $\sim 10^6$ g and a GW spectrum peaked at frequency $(10^5-10^6)$ Hz. More recently, the authors in Ref. \cite{Correa:2022ngq} explain the full abundance of dark matter in the form of PBHs in mass range $(10^{17}-10^{22})$ g from a model of warm natural inflation. Another interesting study considering inflaton dissipation to be effective only for a few efolds, rather than for the full duration (as in warm inflation), was carried out recently in Ref. \cite{Ballesteros:2022hjk}.  By choosing the peak of the primordial power spectrum at some specific scales, PBH in the above mass range are constructed from this model that can explain the full dark matter abundance. Thus, a study of the dissipative effects during inflation leading to small scale features is crucial for understanding the physics of the early Universe and the dark matter.
	
	All the above models of warm inflation considered a minimally coupled inflaton with a canonical kinetic term. 
	In this study, we explore the effects of a non-canonical kinetic term in the evolution of warm inflationary models and consider the warm Higgs-G model for the formation of PBHs and the associated GW spectra. 
	While the motivation of Refs. \cite{Motaharfar:2017dxh,Motaharfar:2018mni} was to explore the parameter space consistent with the large scale CMB observations, we focus on simultaneously explaining the CMB, as well as identifying the parameter space for the required abundance of PBHs as dark matter.  The formation of PBH and secondary GW from non-canonical inflation have also been motivated in the standard cold description, such as in Refs. \cite{Lin:2020goi,Lin:2021vwc,Gao:2020tsa,Yi:2020cut,Teimoori:2021pte,Papanikolaou:2022did}.

	We consider two cases of warm Higgs-G inflation models with a quartic Higgs potential and a dissipation coefficient with a linear dependence on temperature $\Upsilon (\phi, T) \propto T$. For the Galileon term we consider a general form $G(\phi, X) \propto \phi^{2p+1}X^q$  and work with two models characterised by ($p=0$, $q=1$) and ($p=1$, $q=1$).
	The presence of the dissipation term as well as the non-canonical kinetic term damp the inflaton evolution in these models.
	While PBHs could not be produced in this canonical warm inflationary setup, we find that in the non-canonical models in the presence of G-term as considered in our paper, the power spectrum is hugely enhanced at the small scales. As a result, PBHs over a wide mass range can be generated in our scenario, which includes the asteroid mass range $(10^{17}-10^{23}$) g, for which PBHs can constitute the full dark matter abundance. We further calculate the scalar induced GW spectrum associated with these PBHs and find that GW with peak frequency in the sensitivity limits of the forthcoming detectors, such as LISA, BBO, DECIGO, etc. are generated in our models. Thus, future detections or non-detections of GW would be a useful test to these inflationary models. Moreover, we find that these models are also consistent with the swampland and the TCC conjectures and thus, they belong to the viable landscape of UV complete theories.

	This paper is organised as follows: 
	We begin with an introduction to the basics of warm inflation in Section \ref{WI}. Then we describe the inflationary dynamics in warm-G inflation in Section \ref{WGI}. We then introduce our warm Higgs-G inflation model in Section \ref{WHGI}. After parameterizing it in terms of the model parameters and discussing the predictions with CMB observations and swampland conjectures, we discuss the theory of PBH formation and the associated scalar induced GW in Section \ref{PBH}. Then we discuss the results of our model in Section \ref{results} and summarize the paper in Section \ref{summary}.
	
	\section{Basics of Warm Inflation}
	\label{WI}
	In warm inflation, one
	considers the dissipative processes during inflation based on the principles
	of non-equilibrium field theory for interacting quantum fields. 
	The inflaton is assumed to be near-equilibrium and evolving slowly as compared to the microphysics timescales in the adiabatic approximation, see Refs. \cite{Gleiser:1993ea,Berera:1998gx,Berera:2007qm}. 
	The dissipation of the inflaton field to radiation remains active throughout the duration of inflation 
	which naturally ends when either the energy density of the radiation bath becomes larger than that of the inflaton or the slow roll conditions no longer hold.
	
	The equation of motion of the inflaton $\phi$ slowly rolling down a potential $V(\phi)$ during warm inflation is modified due to an additional dissipation term $\Upsilon\dot{\phi}$ arising from the inflaton coupling to other fields and is given as
	\begin{equation}
		\ddot{\phi}+3H\dot{\phi}+\Upsilon\dot{\phi}=-V_{,\phi}.
		\label{phieom}
	\end{equation}
	Here, an overdot denotes the derivatives with respect to cosmic time $t$, $H$ is the Hubble parameter, and the subscript $_{,\phi}$ represents the derivative with respect to $\phi$. 
	The term $\Upsilon (\phi,T)$ is the dissipation coefficient which usually depends on $\phi$ and the temperature of the Universe $T$. One can also rewrite Eq. (\ref{phieom}) in terms of a dissipation parameter $Q \equiv \frac{\Upsilon}{3H} $, as
	\begin{equation}	
		\ddot \phi + 3 H( 1+ Q ) \dot\phi + V_{,\phi}=0.
		\label{inflatoneom}
	\end{equation}
	Since $Q$ is dimensionless, $Q>1$ is usually called the strong dissipative regime while $Q<1$ is referred as the weak dissipative regime of warm inflation.
	Due to the inflaton dissipation,  
	there is an energy transfer from the inflaton to the radiation component, given as 
	\begin{equation}
		\dot\rho_r+4H\rho_r=\Upsilon{\dot\phi}^2~.
		\label{rad}
	\end{equation}
	It is assumed that the radiation thermalizes quickly after being produced, thus,
	$\rho_r=\frac{\pi^2}{30}g_* T^4$,
	where $T$ is the temperature of the thermal bath and $g_{*}$ is the effective number of relativistic degrees of
	freedom present during warm inflation.
	Note that the energy-momentum tensor associated with the inflaton and radiation are not individually conserved while the total energy-momentum tensor of the system remains conserved.
	\subsection{Slow roll approximation}
	To achieve an adequate duration of inflation, the potential of inflaton needs to be sufficiently flat, so that the slow roll conditions are  satisfied. This is described in terms of the slow roll parameters, 
	\begin{equation}
		\epsilon_\phi = 
		\frac{M_{Pl}^2}{2}\,\left(\frac{V_{,\phi}}{V}\right)^2, \hspace{1cm}
		\eta_\phi = 
		{M_{Pl}^2}
		\,\left(\frac{V_{,\phi\phi}}{V}\right)
		\label{coldslow}
	\end{equation}
	where $M_{Pl}=\sqrt{\frac{1}{8\pi G_N}} \simeq 2.44\times10^{18}$ GeV is the reduced Planck mass and $G_N$ is the gravitational constant.
	In addition, in warm inflation, there are other slow roll parameters \cite{Hall:2003zp,Moss:2008yb}
	\begin{equation}
		\beta_\Upsilon = 
		{M_{Pl}^2}
		\,\left(\frac{\Upsilon_{,\phi}\,V_{,\phi}}{\Upsilon\,V}\right),  \hspace{0.5cm}
		b=\frac{T V_{,\phi T}}{V_{,\phi}}~,\hspace{0.5cm} c=\frac{T\Upsilon_{,T}}{\Upsilon}.
		\label{warmslow}
	\end{equation}
	Here the subscript $_{,T}$ represents derivative  with respect to $T$. These additional slow roll parameters are a measure of the field and temperature dependence of the inflaton potential and the dissipation coefficient. The stability analysis of warm inflationary solution leads to the following conditions \cite{Moss:2008yb}
	\begin{align} 
		&\epsilon_\phi \ll 1+Q,\hspace{0.8cm} |\eta_\phi| \ll 1+Q,\hspace{0.8cm} |\beta_\Upsilon| \ll 1+Q, 
		\hspace{0.8cm} 0<b\ll\frac{Q}{1+Q}, \hspace{0.8cm} |c|\leq 4.
		\label{slow_roll}
	\end{align}
	Note that, the conditions on the slow roll parameters $\epsilon_\phi$ and $\eta_\phi$ are weaker than the corresponding
	conditions for cold inflation.
	For large dissipation parameter $Q$, the upper limit on  $\epsilon_\phi$ and $ \eta_\phi$ is increased,
	and therefore, the $\eta$ problem is relaxed in warm inflation \cite{Berera:2004vm}.
	The condition on $b$ implies that warm inflation is only feasible when the thermal corrections to the inflaton potential remain small.

	In the slow roll approximation, we can neglect $\ddot \phi$ in Eq. (\ref{inflatoneom}), which gives
	\begin{equation}
		\dot\phi\approx \frac{-V_{,\phi}}{3H(1+Q)},
		\label{phido}
	\end{equation}
	and since $\dot\rho_r$ is small in Eq. (\ref{rad}), we 
	can approximate $\dot\rho_r\approx 0$ and obtain 
	\begin{equation}
		\rho_r 
		\approx \frac{\Upsilon}{4H} {\dot\phi}^2 =\frac{3}{4} Q {\dot\phi}^2,
		\label{rhor}
	\end{equation}
	which indicates that the radiation energy density is determined by the dissipation parameter, as expected.
	\subsection{Dissipation coefficient}
	The dissipation coefficient $\Upsilon(\phi,T)$ is determined by the  microphysics of the coupled inflaton-radiation system, which includes the channel of the inflaton decay to radiation, its coupling strength with other fields, mass and multiplicities of the radiation fields with which inflaton couples, the temperature of the thermal bath, etc. \cite{Moss:2006gt,BasteroGil:2010pb,BasteroGil:2012cm}. Depending on the form of the interaction Lagrangian, there arise different forms of the dissipation coefficient, as discussed in the literature (see, for instance, Refs. \cite{Berera:1998gx,Bastero-Gil:2018yen,Bastero-Gil:2019gao} for a review on the various studies about the finite temperature quantum field theory calculations of the dissipation coefficient). For our purpose, we will use a dissipation coefficient, which is linearly proportional to the temperature of the thermal bath, i.e.
	$	\Upsilon(\phi,T)\propto T.$
	This particular form arises in a supersymmetric warm inflation model wherein the inflaton undergoes a two-stage decay mechanism via an intermediate field into radiation fields \cite{Moss:2006gt}. In the high temperature limit, when the mass of the intermediate field is smaller than the temperature of the thermal bath, one gets the above expression of the dissipation coefficient. 
	Also, this form of dissipation coefficient arises in the ``Little Higgs" models of electroweak symmetry breaking with Higgs as psuedo-Nambu Goldstone boson of a broken $U(1)$ gauge symmetry \cite{Bastero-Gil:2016qru}. Thermal corrections to the inflaton potential in such models can be controlled to be small, as discussed in Refs. \cite{Hall:2004zr,Bastero-Gil:2016qru,Bastero-Gil:2018yen,Bastero-Gil:2018uep}.  For this work, we do not specify any particular microphysical description for the interaction Lagrangian and choose to work with the form $\Upsilon\propto T$. This form of dissipation coefficient has been previously considered in different warm inflation models in the context of CMB observations \cite{Benetti:2016jhf,Bastero-Gil:2017wwl,Arya:2018sgw}. In this paper, we are interested in the small scale imprints arising from this form of the dissipation coefficient.

	\section{Warm G-inflation}
	\label{WGI}
	In this model, the inflaton also has a non-minimal kinetic term in the Lagrangian, as shown in Eq. (\ref{nonmin}). By the definition of warm inflation, the inflaton couples to radiation fields and dissipates its energy into them. The full action, including the interaction and the radiation Lagrangian is given as,
	\begin{equation}
		\mathcal{S}=\int d^4x \sqrt{-g} \left[\frac{M_{Pl}^2}{2} R +X -V(\phi,T)-G(\phi,X)\Box \phi + \mathcal{L}_r+ \mathcal{L}_{int}\right].
		\label{action1}
	\end{equation}
	As mentioned before,  $X$ 
	is the standard kinetic term, $G(\phi,X)\Box \phi$ is the Galileon-like non-linear kinetic term with $G(\phi,X)$ an arbitrary function of $\phi$, $X$.
	Here 
	$V(\phi,T)$ is the field and temperature dependent inflaton potential, $\mathcal{L}_r$ is the Lagrangian for radiation fields and $\mathcal{L}_{int}$ is the interaction Lagrangian for the inflaton and the fields into which it dissipates its energy. The radiation energy density remains subdominant than that of the inflaton during the inflationary phase.

	By varying the action in Eq. (\ref{action1}) with respect to the metric, one obtains the components of stress-energy momentum tensor for the inflaton as \cite{Kobayashi:2010cm,Motaharfar:2017dxh} 
	\begin{eqnarray}
		\rho_\phi &=& X+V(\phi,T)+ 6H G_{,X} X \dot{\phi} - 2G_{,\phi} X ,
		\label{rhophi}\\
		p_\phi &=& X-V(\phi,T)- 2 (G_{,\phi}+G_{,X} \ddot{\phi})X,
		\label{pressure}
	\end{eqnarray}
	where $G_{,\phi}=\partial{G}/\partial{\phi}$, and $G_{,X}=\partial{G}/\partial{X}$. Also, due to the presence of Galileon-like non-minimal kinetic term, the Klein-Gordon equation for the inflaton gets modified as \cite{Motaharfar:2017dxh},
	\begin{equation}
		\mathcal{B}\ddot{\phi}(t)+3H \mathcal{A}\dot{\phi}(t)+ V_{,\phi}(\phi,T)=0,
		\label{fulleom}
	\end{equation}
	with
	\begin{equation}
		\mathcal{B}=1+6H G_{,X} \dot{\phi} +6H G_{,XX} X \dot{\phi} -2 G_{,\phi} - 2 G_{,X \phi} X,
		\label{Bfull}
	\end{equation}
	and
	\begin{equation}
		\mathcal{A}=1+Q+3H G_{,X} \dot{\phi} +\frac{\dot{H}\dot{\phi}G_{,X}}{H} -2 G_{,\phi} + 2 G_{,X \phi} X -\frac{G_{,\phi\phi}\dot{\phi}}{3H}.
		\label{A}
	\end{equation}
	In the limit, $G\to 0$ in Eq. (\ref{fulleom}), one recovers the warm inflation scenario in a minimal setup, as given in Eq. (\ref{phieom}). From Eq. (\ref{fulleom}),  (\ref{inflatoneom}), 
	we see that both the dissipation and the Galileon interaction terms contribute  as a damping term that further slows down the inflaton evolution.
	However, the radiation energy density evolves in the same way as in Eq. (\ref{rad}).
	
	\subsection{Slow roll conditions}
	In warm G-inflation, apart from the slow roll parameters given in Eqs. (\ref{coldslow}) and (\ref{warmslow}), one also has the following dimensionless parameters \cite{Motaharfar:2017dxh}
	\begin{equation}
		\delta_X=\frac{X}{M_{Pl}^2 H^2}, \hspace{1cm}\delta_{GX}=\frac{\dot{\phi}XG_{,X}}{M_{Pl}^2 H},\hspace{1cm}\delta_{G\phi}=\frac{X G_{,\phi}}{M_{Pl}^2 H^2}.	
	\end{equation}
	The requirement of slow roll during inflation demands the validity of the following conditions, as shown in Ref. \cite{Motaharfar:2017dxh}
	\begin{equation}
		|\epsilon_\phi|,|\eta_\phi|,|\beta_\Upsilon|\ll \mathcal{A}, \hspace{0.6cm}
		|\delta_X|, |\delta_{GX}|,|\delta_{G\phi}|\ll1, \hspace{0.6cm} 0<b\ll\frac{Q}{\mathcal{A}}, \hspace{0.6cm} |c|\leq 4,\hspace{0.6cm} |G_{,\phi}|=\frac{\delta_{G\phi}}{\delta_{X}}\ll1.
	\end{equation}
	
	In the slow roll approximation, $\dot{H}, G_{,\phi}, G_{,X \phi}, G_{\phi\phi}$ terms are small, and thus Eqs. (\ref{Bfull}) and (\ref{A}) can be approximated as
	\begin{equation}
		\mathcal{B}\approx 1+6H G_{,X} \dot{\phi} +6H G_{,XX} X \dot{\phi}.
		\label{B1}
	\end{equation}
	\begin{equation}
		\mathcal{A}\approx 1+Q+3H G_{,X} \dot{\phi}. 
	\end{equation}
	Moreover, the contribution of $\ddot{\phi}$ is negligible in the slow roll approximation and the thermal corrections to the inflaton potential are demanded to be small by construction. Thus Eq. (\ref{fulleom}) can be written as
	\begin{equation}
		3H \mathcal{A}\dot{\phi}(t)+ V_{,\phi}\simeq0.
		\label{phieo1}
	\end{equation} 
	Also, the radiation energy density does not evolve much during the slow roll, thus Eq. (\ref{rad}) is approximated as
	\begin{equation}
		\rho_r\approx\frac{3}{4} Q {\dot\phi}^2.
		\label{radrho}
	\end{equation}
	Next, one can define an effective Galileon dissipation parameter as
	$$Q_G\equiv Q/\mathcal{B}$$ and from Eqs. (\ref{phieo1}) and (\ref{radrho}), obtain 
	\begin{equation}
		3H \mathcal{B}\dot{\phi}\left(Q_G+\frac{\delta_X+3\delta_{GX}}{\delta_X+6(\kappa_X+1)\delta_{GX}}\right)+V_{,\phi}\simeq0
		\label{phiG}
	\end{equation}
	where $\kappa_X=\frac{X G_{,XX}}{G_{,X}}$, and
	\begin{equation}
		\rho_R=C_RT^4\approx\frac{3}{4} \mathcal{B}Q_G {\dot\phi}^2
		\label{rhoR}
	\end{equation}
	where $C_R=\frac{\pi^2}{30}g_*.$
	When the G-term dominates the evolution, i.e., $|\delta_X|\ll|\delta_{GX}|$, then Eq. (\ref{phiG}) becomes
	\begin{equation}
		3H \mathcal{B}\dot{\phi}\left(Q_G+\frac{1}{2(\kappa_X+1)}\right)+V_{,\phi}\simeq0.
		\label{phiGapprox}
	\end{equation}
	This equation governs the dynamics of the inflaton field during warm G-inflation.
	
	\subsection{Primordial Power Spectrum}
	Due to the presence of non-zero temperature during warm inflation, the primordial curvature power spectrum is dominantly sourced by the thermal fluctuations \cite{Hall:2003zp,Graham:2009bf,Ramos:2013nsa}. The inflaton fluctuations are obtained by solving the stochastic Langevin equation sourced by thermal noise in the radiation bath. The intensity of noise depends on the dissipation coefficient through the fluctuation-dissipation theorem \cite{Gleiser:1993ea,Berera:1998gx,Berera:2007qm}. When the dissipation coefficient has a temperature dependence $(\Upsilon\propto T^c$), the radiation fluctuations also couple to the inflaton fluctuations and leads to a growth in the power spectrum for $c>0$ and a suppression when $c<0$ \cite{Graham:2009bf}. 
	Following the calculation for warm inflation of Ref. \cite{Graham:2009bf}, the power spectrum 
	for warm G-inflation model has been calculated in Ref. \cite{ Motaharfar:2018mni}, and is given as\footnote{In this expression, thermal fluctuations are the dominant contributions to the primordial power spectrum.}
	\begin{equation}
		\Delta^2_\mathcal{R}|_{c_s k=aH}=\left(\frac{\sqrt{3}H^3T \sqrt{1+Q_G}}{4\pi\sqrt{\pi}c_s \dot{\phi}^2}\right)\left(1+\frac{Q_G}{Q_c}\right)^{3c}.
		\label{power}
	\end{equation}
	Here the first bracketed term corresponds to the case when there is no coupling of radiation fluctuations to the inflaton fluctuations, i.e. there is no temperature dependence in $\Upsilon$ ($c=0$). The term  $\left(1+\frac{Q_G}{Q_c}\right)^{3c}$ represents the growth factor, and for a fixed $c$ and $Q_G$, it varies inversely to $Q_c$. While if we fix $c$ and $Q_c$, the behaviour of the growth factor is directly proportional to the value of $Q_G$.
	In the above expression, the effective sound speed $c_s$ is obtained from
	\begin{equation}
		c_s^2=\frac{\delta_X+4\delta_{GX}}{\delta_X+6(\kappa_X+1)\delta_{GX}}
		\label{cssquare}
	\end{equation}
	and the factor $Q_c$ is given as
	\begin{equation}
		Q_c=\left(\left[\frac{G^{3,1}_{1,3}\left(\frac{1}{12 c_s^2}|^{\hspace{0.6cm}1-3c/2\hspace{0.2cm}}_{2-c/2,\hspace{0.2cm}0,\hspace{0.2cm}5/2}\right)}{2^{3c}\Gamma_R(\frac{3c}{2})\Gamma_R(\frac{3c}{2}+\frac{5}{2})\Gamma_R(2+c)}\right]\frac{\Gamma_R(3c+\frac{3}{2})}{\Gamma_R(\frac{3}{2})}\right)^{-\frac{1}{3c}}
  \label{Qc}
	\end{equation}
	where $G^{3,1}_{1,3}$ is the Meijer-G function, and $\Gamma_R$ refers to the Gamma function. From this expression, we can see that $Q_c$ is inversely proportional to $c_s$. As $c_s$ decreases, $Q_c$ increases, which in turn makes the growth weaker, and vice-versa.  We also stress the importance of parameter $c$, which is a measure of temperature dependence in the dissipation coefficient, i.e. $\Upsilon\propto T^c$. For positive values, as we increase $c$, the growth factor increases.
	In the G-dominated regime ($|\delta_X|\ll|\delta_{GX}|$), Eq. (\ref{cssquare})  can be approximated as
	\begin{equation}
		c_s^2\approx\frac{2}{3(\kappa_X+1)}=\frac{2 G_{,X}}{3(G_{,X}+XG_{,XX})}.
		\label{csapprox}
	\end{equation} The effects of Galileon-like terms appear in modifying the sound speed and in the limit $G\rightarrow0$,  $c_s^2\rightarrow 1$, one recovers the canonical warm inflation. However, in this study, we are interested in the G-dominated regime and using the expression (\ref {power}) for parameterizing the power spectrum in terms of model parameters.

	The primordial power spectrum could also be damped because of the shear viscous pressure in the radiation \cite{Bastero-Gil:2011rva}. This would lead to a modified expression of the primordial power spectrum with a comparatively slower growth rate or even a total suppression of the growth caused by inflaton and radiation coupled fluctuations. However, in this work, we do not consider any such effects and it would be noteworthy to consider them in future studies.

	\section{Warm Higgs-G model}
	\label{WHGI}
	We will now describe the model we have considered in this study, warm Higgs-G inflation.
	As the name suggests, we have a Galileon-like non-minimal kinetic term to the Standard Model Higgs boson in a warm inflation setup. The action for Higgs-G in warm inflation is given from the Refs. \cite{Kamada:2010qe,Kamada:2013bia,Motaharfar:2018mni} as
	\begin{equation}
		\mathcal{S}_{WHGI}=\int d^4x \sqrt{-g} \left[\frac{M_{Pl}^2}{2} R -|D_\mu \mathcal{H}|^2-\lambda(|\mathcal{H}|^2-v^2)^2-\frac{2\mathcal{H}^\dagger}{M^4} D_\mu D^\mu \mathcal{H} |D_\mu \mathcal{H}|^2+ \mathcal{L}_r+ \mathcal{L}_{int}\right]
		\label{fullaction}
	\end{equation}
	where $D_\mu$ is the covariant derivative with respect to the Standard Model gauge symmetry, $\mathcal{H}$ is the SM Higgs doublet, and $M$ represents some mass parameter. Higgs inflation is driven by the neutral component of Higgs field $\phi$, which has a self-coupling constant $\lambda$ and the vacuum expectation value of Higgs after electroweak symmetry breaking, $v=246$ GeV.  We will consider the scenario when $\phi\gg v$, in which case the simplified action is given by
	\begin{equation}
		\mathcal{S}_{WHGI}=\int d^4x \sqrt{-g} \left[\frac{M_{Pl}^2}{2} R +X -\frac{\lambda\phi^4}{4}-\frac{\phi X}{M^4}\Box \phi + \mathcal{L}_r+ \mathcal{L}_{int}\right],
		\label{action}
	\end{equation}
	where we have approximated the Higgs potential as quartic potential,
	\begin{equation}
		V(\phi)\simeq \frac{\lambda\phi^4}{4}.
	\end{equation}
	In this study, we would choose a general form of non-linear Galileon interaction term ($G(\phi,X)\Box \phi$ term in the Lagrangian) as \cite{Kamada:2013bia,Motaharfar:2017dxh,Deffayet:2010qz}
	\begin{equation}
		G(\phi,X)= -\frac{\phi^{2p+1} X^q}{M^{4q+2p}}
	\end{equation}
	where $p$ and $q$ are some positive integers. We consider two models with G-term as $-\frac{\phi X}{M^4}$ (for $p=0,q=1$) and $-\frac{\phi^3 X}{M^6}$ (for $p=1,q=1$). For the warm inflation setup, we choose the interaction Lagrangian in inflaton-radiation system such that the dissipation coefficient is given by \cite{Moss:2006gt,BasteroGil:2010pb,Bastero-Gil:2016qru}
	\begin{equation}
		\Upsilon(\phi,T)= C_T T.
		\label{ups}
	\end{equation}
	As discussed before, this form of the dissipation coefficient is obtained in certain microphysical descriptions of warm inflation. The temperature dependence in the dissipation coefficient couples the inflaton fluctuations to the fluctuations in the radiation fields and leads to a growth function in the primordial power spectrum. Other well-motivated forms of temperature and field dependence in the dissipation coefficient are equally interesting to explore further.
	
	\subsection{Evolution equations in warm Higgs-G inflation}
 \label{whgi}
	Now, we will study the inflationary dynamics for our warm Higgs-G inflation model, focussing on the regime wherein the G-term dominates the inflaton evolution. We begin with parameterizing the primordial power spectrum in Eq. ({\ref{power}}) in terms of model parameters.  First, we write the Friedmann equation for our model
	when the inflaton potential dominates the energy density of the Universe, as 
	\begin{equation}
		H^2\simeq\frac{\text{$\lambda$}}{12}  \text{$M_{Pl}$}^2 \left(\frac{\phi}{M_{Pl}}\right)^4.
		\label{Hubble}
	\end{equation}
	Next, we want to evaluate $\dot\phi$. For this, 
	we simplify Eq. (\ref{B1}) for our model and obtain
	\begin{equation}
		\mathcal{B}=-6Hq^2\dot{\phi}~\frac{\phi^{2p+1} X^{q-1}}{M^{4q+2p}}
		\label{B}
	\end{equation}
	where $X=-\frac{\dot{\phi}^2}{2}.$
	Then by substituting the expression for $\mathcal{B}$ in Eq. (\ref{phiGapprox}), we get
	\begin{equation}
		\dot{\phi}=-M_{Pl}^2 \zeta_1^\frac{1}{2q}\left(Q_G+\frac{1}{2q}\right)^{-\frac{1}{2q}} \left(\frac{\phi}{M_{Pl}}\right)^{-\frac{(p+1)}{q}},
		\label{phidot}
	\end{equation}
	where $\zeta_1=\frac{2^q}{3 q^2} \left(\frac{M}{\text{$M_{Pl}$}}\right)^{2 p+4 q}$. 
	We would further express the field evolution in terms of the parameter $Q_G$.  From the form of dissipation coefficient in Eq. (\ref{ups}), and the definitions of $\Upsilon, Q, Q_G$, we can write $T= \frac{3H\mathcal{B} Q_G}{C_T}$, and then substitute it in Eq. (\ref{rhoR}). Then using  Eq. (\ref{phidot}), we get a relation between $\phi$ and $Q_G$ as
	\begin{equation}
		\frac{\phi}{M_{Pl}}=\left[\frac{\text{$Q_G^3$} \left(\frac{1}{2 q}+\text{$Q_G$}\right)^{\frac{5-6 q}{2 q}}}{\text{$\zeta_2$}}\right]^{-\frac{q}{5 p+11 q+5}}
		\label{phibyMPl}
	\end{equation}
 where 
	$\text{$\zeta_2$}=\frac{\sqrt{3} \text{$C_T^4$}}{2 \text{$C_R$} \text{$\lambda$}^{7/2}} \text{$\zeta_1$}^{\frac{5}{2 q}}$.
	Later, we also plot and discuss the total field excursion for different values of $Q_G, p$ and $q$ in our warm inflation models in the context of swampland conjectures.
	
	We next evaluate the expression for the temperature of the thermal bath of radiation.
	Using  Eqs. (\ref{rhoR}), (\ref{B}) and (\ref{phidot}), we get
	\begin{equation}
		\frac{T}{M_{Pl}}= Q_G^{\frac{1-3 a}{4}} \left(\frac{1}{2 q}+Q_G\right)^{\frac{2q(3 a-1) -5 a-1}{8 q}} \left(\frac{\sqrt{3\lambda}}{2 C_R}\zeta_1^{\frac{1}{2q}}\zeta_2^a\right)^{\frac{1}{4}}
		\label{TbyMPl}
	\end{equation}
	where 
	$a=\frac{-p+q-1}{5 p+11 q+5}$.
	Finally, with the Eqs. (\ref{Hubble}), (\ref{phidot}),  (\ref{phibyMPl}) and (\ref{TbyMPl}), we have parameterized the primordial power spectrum in terms of parameters $Q_G$, $\lambda$, $C_T$, $M$,  $p, q$, and $c_s$ (which is a function of $p,q$). For a chosen model i.e. fixed $p, q$ (and $c_s$ accordingly), and a fixed $Q_G$ value at the pivot scale, we have three unknown parameters, $\lambda, C_T,$ and $ M$. The variable $Q_G$ is dynamical during inflation. 
	We know that the end of warm inflation is determined from the equation $\epsilon_H=-\frac{\dot{H}}{H^2}=1$, which gives 
	\begin{equation}
		\frac{1}{2 \zeta_3}	(Q_G^{end})^{-3b}\left(\frac{1}{2 q}+Q_G^{end}\right)^\frac{(6q-5)b+1}{2q}=1.
		\label{QGend}
	\end{equation}
	where $
	\text{$\zeta_3$}=2 \sqrt{\frac{3}{\text{$\lambda$}}} \text{$\zeta_1$}^{\frac{1}{2 q}} \text{$\zeta_2$}^{-b}
	$ and $Q_G^{end}$ is the value of $Q_G$ at the end of inflation. Thus, for a warm inflation model with a fixed $\lambda$, the above equation gives a relation between $Q_G^{end}$, $C_T$ and $M$. But, since we also have $Q_G^{end}$ as unknown, we need to evaluate the evolution of $Q_G$ as a function of number of efolds $N_e$ for our warm inflation model. 
	In our notation, we count the number of efolds from the end of inflation, i.e. $N_{end}=0$. From Eq. (\ref{phibyMPl}), we have a relation between $\phi$ and $Q_G$. On differentiating both sides with $N_e$ and using the fact that $d N_e=-H dt$, we have $d\phi/dN_e=-\dot{\phi}/H$. On substituting for $\dot{\phi}$ from Eq. (\ref{phidot}), we get
	\begin{equation}
		\frac{dQ_G}{dN_e}=-\zeta_3 (22q + 10p + 10)\frac{Q_G^{3b+1}}{5Q_G+3}\left(\frac{1}{2 q}+Q_G\right)^{-\frac{2q(3b-1) -5 b+1}{2 q}}
		\label{dQGbydN}
	\end{equation}
	where $	b=\frac{p+3 q+1}{5 p+11 q+5}.$
	We next integrate this equation from the pivot scale till the end of inflation to obtain $Q_G^{end}$, as
	\begin{equation}
		\zeta_3 N_e = F(Q_G^{end})-F(Q_G^{P})
		\label{c2}
	\end{equation}
	where $$F(Q_G)=-\frac{q Q_G^{-3b}\left(\frac{1}{2 q}+Q_G\right)^{\frac{6bq -5 b+1}{2 q}}\left(5-\frac{_2F_1(1,\frac{1-5b}{2q};1-3b;-2q Q_G)}{b}\right)}{(5b-1)(5p+11q+5)}.$$
	With this relation, for a fixed  number of efolds of inflation $N_e=50$ (or $60$) and a chosen value of $\lambda$ and $Q_G$ at the pivot scale ($Q_G^{P}$), we again obtain the $Q_G$ value at the end of inflation ($Q_G^{end}$) as a function 
	of $C_T$ and $M$. Since we have three variables ($Q_G^{end}, C_T, M$) and two equations (\ref{QGend}), (\ref{c2}), we need to provide an additional constraint to know these exactly.
	We use the normalisation of primordial power spectrum at the pivot scale ($k_P=0.05$ Mpc$^{-1}$) as $\Delta^2_\mathcal{R}(k_P)=2.1\times10^{-9}$ and therefore obtain the values of all the variables. 
	In this way, we parameterize the power spectrum for our warm Higgs-G model in terms of model parameters and estimate their values compatible with the CMB observations. We also list the values of all the parameters used in our analysis in Table \ref{tab:1} and \ref{tab:2}. The background evolution of inflaton field, temperature of the radiation bath $T$, dissipation parameter $Q$, and energy densities in inflaton and radiation are shown in previous study in the context of CMB observations \cite{Motaharfar:2018mni}. Here we focus on the small scale imprints of these models in the formation of PBH and secondary gravitational waves. 
	
	\subsection{Constraints on warm Higgs-G inflation model from CMB}
	\label{Qprange}
\begin{figure}[]
		\subfigure[]{
			\label{fig:1a}
			\includegraphics[scale=0.8]{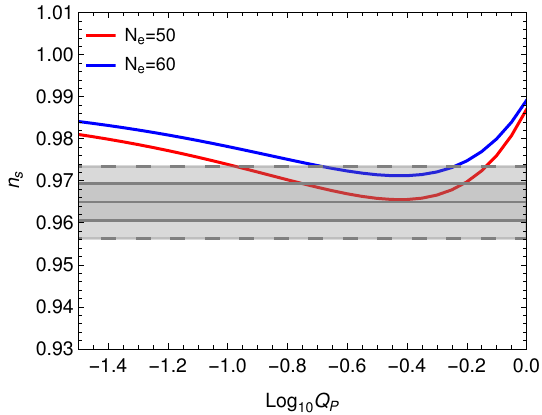}}
		\hspace{0.2cm}
		\subfigure[]{
			\label{fig:1b}
			\includegraphics[scale=0.8]{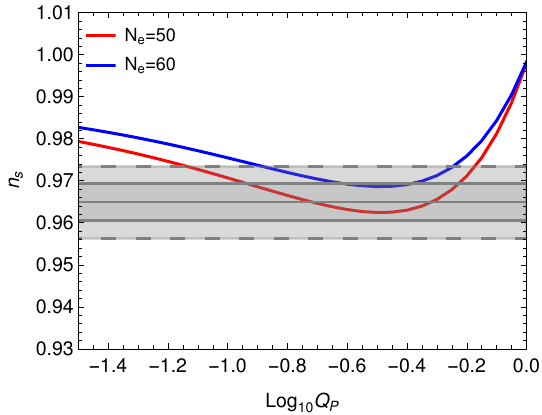}}	
		\caption{Plots of the spectral index $n_s$ vs. $Q_P$ (equal to the value of $Q_G$  at the pivot scale) for 50 (red) and 60 (blue) efolds of inflation for \ref{fig:1a} Model I: $p=0, q=1$, and \ref{fig:1b} Model II: $p=1, q=1$. The grey shaded bands represent the $1\sigma$ and $2\sigma$ allowed ranges from the Planck observations.}
		\label{ns} 
	\end{figure}

We first identify the range of model parameters consistent with the large scale CMB observations. For this, we plot the $n_s$ predictions of our models and compare it with the Planck observations in Fig. \ref{ns}.
 We choose $\lambda=0.13$ (the largest value of Higgs self-coupling from quantum field theory) for our study. In all the plots, for different values of $Q_G$ (chosen at the pivot scale), we have a different set of $C_T$ and $M$ values that satisfy the normalisation condition for the primordial power, which are listed in Table \ref{tab:1} and \ref{tab:2}. We consider the following set of values for $p$ and $q$.\\

 \textbf{Model I:} $p=0, q=1$, which effectively represents the Galileon interaction term, $G=-\frac{\phi X}{M^4}$. For these values of $p,q$, the sound speed corresponds to $c_s=\sqrt{\frac{2}{3}}=0.816$ and the factor in Eq. (\ref{Qc}), $Q_c =8.7476$. Also, the values of parameters $C_T \sim 10^{12}$ and $M\sim 10^{-2}$ GeV. With this set of parameters, the condition of G-dominated regime $|\delta_X|\ll|\delta_{GX}|$ is also fulfilled.
 We plot the spectral index $n_s$ versus $Q_G$ value at the pivot scale in Fig. \ref{fig:1a} for $50$ and $60$ efolds of inflation. The coloured band in the Figure represents the Planck $2\sigma$ allowed range for the $n_s$. We obtain that for $N_e=50$, $10^{-0.98}\leq Q_G\leq 10^{-0.135}$ is consistent within the $n_s- 2\sigma$ bounds, while for $N_e=60$, the allowed range is $10^{-0.68}\leq Q_G\leq10^{-0.25}$. Note that although the value of $Q_G (=\frac{Q}{\mathcal{B}})$ parameter at the pivot scale is small, the factors $\mathcal{B}$ and $Q$ are very large, and both Galileon and dissipative effects combine to reduce the Hubble value. 
   \begin{table}[]
   \centering
		\begin{tabular}{|c|c|c|c|c|}\hline
			& \multicolumn{4}{c|}{$p=0, q=1$} \\ \hline
			& \multicolumn{2}{|c|}{$N_e=50$} 
			& \multicolumn{2}{|c|}{$N_e=60$} \\ \hline	
			$Q_G$ &	$10^{-0.135}$ &  $10^{-0.98}$ & $10^{-0.25}$ & $10^{-0.68}$ \\ \hline
			$M (GeV)$ &	$0.066$ & $0.027$ & $ 0.033 $  & $0.02$ \\ \hline
			$C_T$ &	$9.7\times10^{11}$& $9.4\times10^{11} $ &  $1.6\times10^{12}$ &  $1.5\times10^{12}$ \\ \hline
			$\phi/M_{Pl}$ &	$7.3\times10^{-10}$  & $5.9\times10^{-10}$  &  $5.6\times10^{-10}$ & $5.1\times10^{-10}$   \\ \hline
			$\dot{\phi}$ &	$-4.4\times10^{6}$  & $-1.3\times10^{6}$  &  $-1.5\times10^{6}$ & $-9.3\times10^{5}$  \\ \hline	
			$T/M_{Pl}$ &		$4.2\times10^{-11}$  & $2.2\times10^{-11}$  &  $2.9 \times 10^{-11}  $ & $2.2 \times 10^{-11}  $ \\ \hline	
			$T/H$ & $7.5\times10^{8}$  & $6\times10^{8}$  &  $9 \times 10^8  $ & $7.9 \times 10^8  $  \\ \hline	
			$H (GeV)$ &	$0.13$  & $0.088$  &  $0.08 $ & $0.067$ \\ \hline	
			$\mathcal{B}$ &	$3.3\times10^{21}$  & $1.8\times10^{21}$  &  $8.3 \times 10^{20}  $  & $1.9\times10^{21}$ \\ \hline
			$Q$ &	$2.4\times10^{20}$  & $1.87\times10^{20}$  &  $4.7 \times 10^{20}  $ & $4.0\times10^{20}$   \\ \hline	
		\end{tabular}
			\caption{List of all the parameter values at the pivot scale for Model I ($p=0, q=1$), obtained by solving the evolution equations in Section \ref{whgi}. The value of $\lambda=0.13$ and the normalisation of power spectrum is considered as $\Delta^2_\mathcal{R}(k_P)=2.1\times10^{-9}$.}
			\label{tab:1}
	\end{table}  

 	\begin{table}[]
	\centering
		\begin{tabular}{|c|c|c|c|c|}\hline
			 & \multicolumn{4}{c|}{$p=1, q=1$} \\ \hline
			 & \multicolumn{2}{|c|}{$N_e=50$} 
			& \multicolumn{2}{|c|}{$N_e=60$} \\ \hline	
			$Q_G$ & $10^{-0.17}$ &  $10^{-1.14}$   & $10^{-0.24}$ & $10^{-0.88}$ \\ \hline
			$M (GeV)$ & $194.7$ & $69.5$ & $110.5$ & $58.7$\\ \hline
			$C_T$ & $9.8\times10^{11}$ & $1.1\times10^{12}$ & $1.6\times10^{12}$ & $1.8\times10^{11}$\\ \hline
			$\phi/M_{Pl}$  & $7.3\times10^{-10}$& $4.8\times10^{-10}$& $5.5\times10^{-10}$ & $4.3\times10^{-10}$  \\ \hline
			$\dot{\phi}$ & $-4.3\times10^{6}$& $-6.3\times10^{5}$& $-1.4\times10^{6}$& $-4.5\times10^{5}$   \\ \hline	
			$T/M_{Pl}$ & $4.1 \times 10^{-11}  $& $1.6 \times 10^{-11}  $& $2.9 \times 10^{-11}  $& $1.6 \times 10^{-11}  $ \\ \hline	
			$T/H$  &$7.4 \times 10^8  $ &  $6.4 \times 10^8  $&  $9.2 \times 10^8  $&  $8.2\times 10^8  $ \\ \hline	
			$H (GeV)$  & $0.13$ & $0.06$ & $0.076$&$0.047$\\ \hline	
			$\mathcal{B}$ & $3.6\times10^{20}$& $3.3\times10^{21}$& $8.6\times10^{20}$& $3.7\times10^{21}$ \\ \hline
			$Q$ &	$2.4\times10^{20}$ &	$2.4\times10^{20}$ &	$5\times10^{20}$ &	$4.9\times10^{20}$ \\ \hline	
		\end{tabular}
		\caption{List of all the parameter values for Model II ($p=1, q=1$).}
		\label{tab:2}
	\end{table}
 
	\textbf{Model II:} $p=1, q=1$, which effectively represents the Galileon interaction term, $G=-\frac{\phi^3 X}{M^6}$. For these values of $p, q$ also, the sound speed  $c_s=\sqrt{\frac{2}{3}}$ and the factor $Q_c =8.7476$. The parameters $C_T\sim 10^{12}$, $M\sim10^2$ GeV and $|\delta_X|\ll|\delta_{GX}|$ in this model, thereby implying a G-dominated regime. To estimate the range of $Q_G$ values consistent with the CMB, we plot the $n_s$ versus $Q_G$ value at the pivot scale for $50$ and $60$ efolds of inflation in Fig. \ref{fig:1b}. We obtain that for $N_e=50$, $10^{-1.14}\leq Q_G\leq 10^{-0.17}$ is consistent within the $n_s- 2\sigma$ bounds, and for $N_e=60$, the allowed range is $10^{-0.88}\leq Q_G\leq10^{-0.24}$. 

We see from the Fig. \ref{ns} that the spectrum is red-tilted ($n_s<1$) for small $Q_G$ values, while it becomes blue-tilted ($n_s>1$)  as $Q_G$ increases. Following Eq. (\ref{dQGbydN}), we can see that in our models, the dissipation parameter increases as inflation proceeds. Thus, the spectral index can even become highly blue-tilted for the small-scale modes, which could be problematic, as the amplitude of the primordial power spectrum shoots above the perturbative regime. This raises a serious issue on the validity of models with a linear dissipation coefficient ($\Upsilon\propto T$). In order to avoid it, we need some mechanism to control the growth of the power spectrum. We will discuss one such example in Section \ref{PPS}.
We stress here that the parameter space obtained in this section is model-specific and it is necessary to check the consistency with the CMB 
	observations, before studying the small-scale evolution and the formation of primordial black holes. 
	\subsection{Swampland conjectures and our model}
	As inflation is described by a low-energy effective field theory, it has to obey some
	criteria, such as the swampland and trans-Planckian censorship conjectures (TCC), in order to embed it in a UV complete
	theory, as follows \cite{Obied:2018sgi,Kehagias:2018uem,Kinney:2018nny,Bedroya:2019snp}:
	\begin{itemize}
		\item \textbf{Distance conjecture:} This criteria puts  a limit on the scalar field range traversed during inflation, and is given as
		\begin{equation}
			\frac{|\Delta \phi|}{M_{Pl}}\leq a
		\end{equation}
		where the constant $a\sim \mathcal{O}(1)$. This implies that small field models of cold inflation are more favourable than the large field models. 
		\item \textbf{de Sitter conjecture:} This criteria gives a lower limit on the slope of inflationary potential as
		\begin{equation}
			M_{Pl}\frac{|V_{,\phi}|}{V}\geq b
		\end{equation}
		where the constant $b\sim \mathcal{O}(1)$. This condition implies that the steep potentials are favourable for inflationary dynamics, which is not actually supported in the standard cold inflation in a minimal setup.
		\item\textbf{ Trans-Planckian Censorship conjecture (TCC):} This criterion demands that the super-Planckian quantum fluctuations never become superhorizon during inflation, which sets a limit on the scale of inflation as \cite{Bedroya:2019tba}
		\begin{equation}
			V^{1/4}<3\times10^{-10} M_{Pl}.
		\end{equation}
		This implies that the energy scale of inflation has to be lower than $\sim 10^9$ GeV to satisfy TCC bound. 
		Such a low energy scale inflation yields a negligible amplitude of primordial gravitational waves and tensor-to-scalar ratio, $r<2.7\times10^{-31}$. 
	\end{itemize}
	It is extremely challenging to satisfy these conjectures in
	a single-field 
	cold inflation model with a canonical kinetic term and a
	Bunch Davies vacuum, to embed them in a UV complete theory. However, if one extends to multifield inflation \cite{Achucarro:2018vey} or curvaton models or by choosing different initial vacuum state \cite{Ashoorioon:2018sqb} or a non-canonical kinetic term, one can construct inflationary models compatible with the swampland distance and de Sitter conjectures \cite{Das:2018hqy,Kehagias:2018uem}. 
	
	\begin{figure}[]
		\subfigure[]{%
			\label{fig:0a}%
			\includegraphics[scale=0.8]{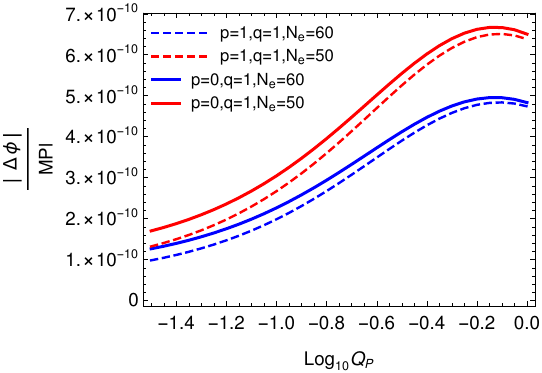}}
		\hspace{0.2cm}
		\subfigure[]{%
			\label{fig:0b}%
			\includegraphics[scale=0.8]{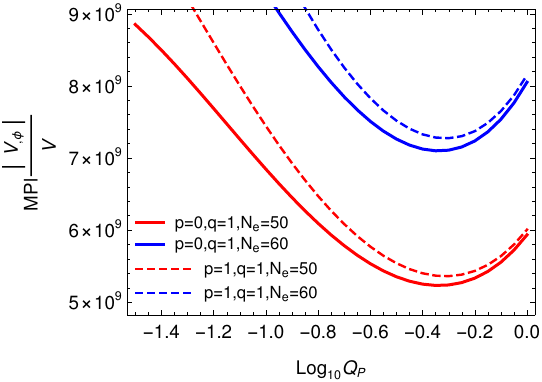}} 
		\caption{Plots of the field excursion $|\Delta \phi|$ (in units of $M_{Pl}$) and the slope of inflationary potentials $M_{Pl} \frac{|V_{,\phi}|}{V}$ vs. $Q_P$ for our warm inflation models. The solid (dashed) red and blue lines represent $50$ and $60$ efolds of inflation with Model I: $p=0,q=1$ (Model II: $p=1,q=1)$.}
		\label{swamp1}
	\end{figure}

	\begin{figure}[]
		\subfigure[]{%
			\label{fig:0c}%
			\includegraphics[scale=0.85]{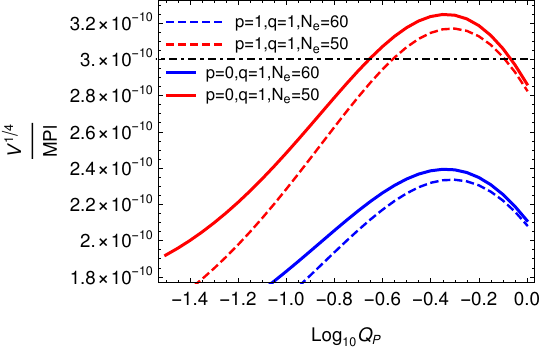}}%
		\hspace{0.2cm}
		\subfigure[]{%
			\label{fig:0d}%
			\includegraphics[scale=0.8]{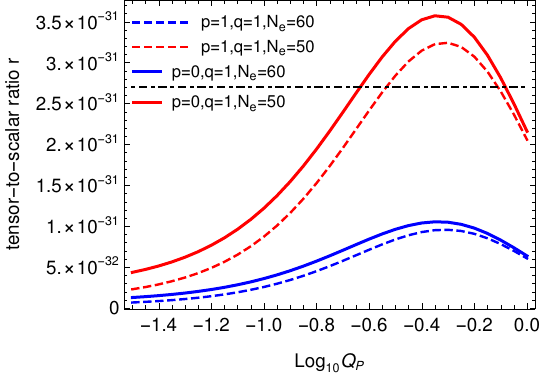}} 
		\caption{Plots of the scale of inflation $V^{1/4}$ (in units of $M_{Pl}$) and the tensor-to-scalar ratio $r$ vs. $Q_P$ for our warm inflation models. The solid (dashed) red and blue lines represent $50$ and $60$ efolds of inflation with Model I: $p=0,q=1$ (Model II: $p=1,q=1)$. The black dot-dashed lines represent the upper bound on both the quantities. }
		\label{swamp2}
	\end{figure}
	Contrary to cold inflation, warm inflation
	is an interesting 
	framework in which these conjectures could be satisfied due to the modified dynamics of inflaton field. There exists a few warm inflationary studies \cite{Das:2018rpg,Motaharfar:2018zyb,Kamali:2019xnt,Das:2019acf,Das:2020xmh,Bertolami:2022knd,Santos:2022exm} in this context, which show that with a strong dissipation, the energy scale of inflation is sufficiently lowered, satisfying the above conjectures. Here we discuss the status of these conjectures for our warm Higgs-G model. In Fig. \ref{swamp1}, we plot the field excursion $|\Delta\phi|/M_{Pl}$ and the slope of inflationary potential ${M_{Pl}}|V_{,\phi}|/V$ for our models with different $Q_G$ value at the pivot scale (denoted by $Q_P$). The solid (dashed) red and blue lines represent Model I (Model II)
	for $50$ and $60$ efolds of inflation, respectively. It is evident from the figure that the swampland distance and de-Sitter conjectures are in agreement with our models, implying that these models lie in the UV complete landscape theories of inflation. Further, we plot the scale of inflation $V^{1/4}/M_{Pl}$ and the tensor-to-scalar ratio $r$ for our warm inflation models in Fig. \ref{swamp2}.  We find that both the TCC and the constraint on the tensor-to-scalar ratio are satisfied for these models with $60$ efolds of inflation, however for $50$ efolds, there is some parameter range of $Q_P$, which does not satisfactorily obey the bound. We again stress that both the dissipative and Galileon effects lead to a suppression of the energy scale of inflation in our models.
	
	To conclude, we have seen that for some parameter space of our warm inflation models, the CMB constraints on $n_s-r$ are well obeyed. Further, the swampland and TCC conjectures are also in agreement with these models, implying that they lie in the viable landscape theories of inflation.  We next study the small scale features of these models in the context of formation of PBH.
	\section{PBH formation and scalar induced gravitational waves}
	\label{PBH}

	In the previous Section, we found that in our models, the primordial power spectrum is turning blue-tilted on the small scales and therefore can generate a significant abundance of PBHs, if the amplitude of the fluctuations is large.
	Further, these overdense perturbations also source the tensor perturbations and lead to secondary gravitational waves. In this Section, we will briefly discuss the PBH formation and the associated GW generation. For more detailed derivation, see Refs. \cite{Ananda:2006af,Baumann:2007zm,Kohri:2018awv,Arya:2019wck,Arya:2022xzc}.
	
	\subsection{PBH formation}
	As mentioned before, PBHs can be produced via the collapse of large inhomogeneities generated during inflation. We consider that the PBHs are generated in the radiation-dominated era when fluctuations reenter the horizon. The mass of the generated PBHs is a fraction, $\gamma$ of the horizon mass at that epoch and is given as \cite{Green:1997sz}
	\begin{equation}
		M_{PBH}(k)=\gamma \frac{4\pi}{3}\rho \left.H^{-3}\right|_{k=aH}.
		\label{MPBH}
	\end{equation}
	where $\rho$, $a$, and $H$ represent the energy density, scale factor and Hubble expansion rate at the time of PBH formation, respectively. 
	Also, as different comoving fluctuation  modes $k$ reenter the horizon at different epochs, the mass of the generated PBHs are given as,
	\begin{align}
		M_{PBH}(k)
		\simeq 5\times 10^{15}{\rm{g}}\left(\frac{g_{*0}}{g_{*i}}\right)^{1/6}\left(\frac{10^{15}{\rm{Mpc}}^{-1}}{k}\right)^{2}
		\label{Mpbh}
	\end{align}
	where $\Omega_{r0}$ is the present day radiation energy density fraction, $g_{*0}$ and $g_{*i}$ represent the relativistic degrees of freedom present in the Universe today and at the time of PBH formation, respectively. The fraction of the horizon mass collapsing into the PBHs is taken as $\gamma=0.2$ \cite{Carr:1975qj}. From Eq. (\ref{Mpbh}), we see that $M_{PBH}\propto k^{-2}$, which suggests that large (small) $k$ leads to small (large) mass PBHs. 
	Further, an important quantity, known as the initial mass fraction of PBHs, $\beta(M_{PBH})$ is defined as
	\begin{equation}
		\beta(M_{PBH})=\frac{\rho_{{PBH},i}}{\rho_{total,i}}
	\end{equation}
	where, $\rho_{{PBH},i}$ and $\rho_{total,i}$ are the energy density of PBHs and the total energy density of the Universe at the time of PBH formation.  Assuming that PBHs are formed in the radiation-dominated era, 
	the expression of $\beta(M_{PBH})$ can be expressed in the form of present-day observables as
	\begin{align}
		\beta({M_{PBH}})=&\frac{\Omega_{PBH,0}({M_{PBH}})}{\Omega_{r0}^{3/4}} \left(\frac{g_{*i}}{g_{*0}}\right)^{1/4} \left(\frac{M_{PBH}}{M_0}\right)^{1/2}\gamma^{-1/2}~~.
		\label{bet}
	\end{align}
	Here, $\Omega_{PBH,0}({M_{PBH}})=\rho_{{PBH},0}/\rho_{{crit},0}$ is the present day PBH energy density fraction, with $\rho_{{PBH},0}$ 
	and $\rho_{{crit},0}$ representing 
	the present energy density of PBHs 
	and the critical energy density of the Universe, respectively. 
	
	The abundance of PBHs is theoretically obtained by the Press-Schechter theory \cite{Press:1973iz}. For an overdense fluctuation reentering the horizon during a radiation-dominated era with an amplitude above a critical value $\delta_c$, the  initial mass fraction of PBHs with mass $M_{\rm{PBH}}$ is 
	given as  \cite{Press:1973iz},
	\begin{align}
		\beta({M_{PBH}})=&\frac{2}{\sqrt{2\pi}\sigma(R)}\int_{\delta_c}^{1} \exp\left( \frac{-\delta^2(R)}{2\sigma^2(R)}\right)d\delta(R)=\mathrm{Erfc}\left(\frac{\delta_c}{\sqrt{2}\sigma(R)}\right)
		\label{betaM}
	\end{align}
	where $\mathrm{Erfc}$ is the  complementary error function, and $\sigma(R)$ is the mass variance evaluated at the horizon crossing defined as,
	\begin{equation}
		\sigma^2(R)=\int_{0}^{\infty}\tilde{W}^2(kR) P_\delta(k)\frac{dk}{k}
		\label{variance}
	\end{equation}
	where $P_\delta(k)$ is the matter power spectrum, and $\tilde{W}(kR)$ is the Fourier transform of the window function.  From Eq. (\ref{bet}), we see that the initial mass fraction of PBH is highly sensitive to the value of $\delta_c$. Therefore, any uncertainty in $\delta_c$ would lead to a large discrepancy in $\beta$. For discussion on this, see review \cite{Carr:2020gox}.
	In our analysis, we assume $\delta_{c}=0.5$, and window function as a Gaussian, i.e., $\tilde{W}(kR)=\exp(-k^2R^2/2)$. The primordial curvature power spectrum $\Delta^2_\mathcal{R} (k)$ for the fluctuations generated during the inflation can be related to the density power spectrum $P_\delta(k)$ as 
	\begin{equation}
		P_\delta(k)=\frac{4(1+w)^2}{(5+3w)^2}\left(\frac{k}{aH}\right)^4 \Delta^2_\mathcal{R} (k),
		\label{PdR}
	\end{equation}
	where $w$ is the equation of the state of the fluid, equal to 1/3 for a radiation-dominated era. Substituting Eq. (\ref{variance}) and (\ref{PdR}) in (\ref{betaM}), we obtain the theoretical estimate of the initial mass fraction for PBHs from our model.

	The abundance of PBHs is bounded from above through different observations, e.g. for the evaporating PBHs, the bounds arise from consequences of evaporation and for non-evaporating PBHs, the upper bound on $\beta$ comes from gravitational lensing, dynamical effects on interaction with astrophysical objects, gravitational wave merger rates, etc. \cite{Green:1997sz,Carr:2009jm,Josan:2009qn,Sato-Polito:2019hws}. 
	This, in turn, then provides an upper limit on the primordial power spectrum. It is seen that for a significant PBH abundance, to have any measurable consequences, we require $\Delta^2_\mathcal{R} (k)\sim \mathcal{O} (10^{-2})$. Thus, we require a blue-tilted power spectrum with an amplitude of this order for significant PBH generation.
	
	\subsection{PBH as dark matter candidate}
	Furthermore, primordial black holes, which have not evaporated completely by now, can constitute the present dark matter abundance. We assume a monochromatic mass function for PBHs. The fraction of dark matter in the PBHs  is defined as,
	\begin{equation}
		f_{PBH}(M_{PBH})=\frac{\Omega_{PBH,0}(M_{PBH})}{\Omega_{CDM,0}}
	\end{equation}
	where $\Omega_{CDM,0}=\rho_{{CDM},0}/\rho_{{crit},0}$ is the fractional cold dark matter (CDM) density at the present, with $\rho_{{CDM},0}$ representing the present energy density of the CDM.
	From Eq. (\ref{bet}), we can substitute for $\Omega_{PBH,0}(M_{PBH})$ and get
	\begin{equation}
		f_{PBH}(M_{PBH})=
		\beta({M_{PBH}})\frac{\Omega_{r0}^{3/4}}{{\Omega_{CDM,0}}}\left(\frac{g_{*i}}{g_{*0}}\right)^{-1/4}\left(\frac{M_{PBH}}{M_0}\right)^{-1/2}\gamma^{1/2}
		\label{fPBH}
	\end{equation}
	where  $M_0=\frac{4\pi}{3}\rho_{crit,0}~H_0^{-3}\approx 4.62\times 10^{22} M_{\odot}$ is the present horizon mass. This expression gives the fractional abundance of primordial black holes in the form of dark matter.  
	As can be seen from Eq. (\ref{fPBH}), for any mass PBH, the fraction in DM is directly proportional to its initial mass fraction. 
	Thus, a bound on the initial abundance limits the fraction of PBHs in the dark matter. However, for a mass window ($10^{17}-10^{23}$) g, the bounds on the initial mass fraction or the fraction of dark matter are not very restrictive.
	For related discussion, see Ref. \cite{Montero-Camacho:2019jte}. 
	Thus, there is a possibility that the PBHs in the asteroid mass range ($10^{17}-10^{23}$) g can constitute the full abundance of dark matter. Hence, PBHs are interesting from the viewpoint of dark matter phenomenology.

	\subsection{Scalar induced gravitational waves}
	Now we discuss the gravitational wave spectrum associated with the primordial black holes. 
	As we have discussed above, for PBH production, the amplitude of primordial fluctuations is enhanced at small scales. Thus, 
	these overdense scalar modes can source the tensor modes at the second order of cosmological perturbation theory 
	and inevitably lead to the generation of scalar induced gravitational waves (SIGW) spectrum \cite{Mollerach:2003nq,Ananda:2006af,Baumann:2007zm,Espinosa:2018eve,Kohri:2018awv}. For a recent review, see Ref. \cite{Domenech:2021ztg}.
	
	The gravitational wave energy density parameter per logarithmic $k$ interval is given by \cite{Baumann:2007zm,Kohri:2018awv}:
	\begin{align}
		\Omega_{\mathrm{GW}}(\eta,k) =\frac{\rho_{\mathrm{GW}}(\eta,\mathrm{k})}{\rho_{\mathrm{tot}}(\eta)}=\frac{1}{24}\left( \frac{\mathrm{k}}{a(\eta)H(\eta)}\right)^{2} \overline{P_h(\eta, \mathrm{k})}
		\label{eq:omegagw}
	\end{align}
	where $\rho_{\mathrm{tot}}(\eta)$ is the total energy density and $\overline{P_h(\eta, \mathrm{k})}$ is the dimensionless tensor power spectrum averaged over time given by
	\cite{Baumann:2007zm,Kohri:2018awv}
	\begin{equation}
		\overline{\mathcal{P}_{h}(\eta,k)} = 4\int\limits_{0}^{\infty} dv \int\limits_{|1-v|}^{|1+v|} du \left[ \frac{4v^{2}-(1+v^{2}-u^{2})^{2}}{4vu}\right]^{2} \overline{\mathrm{I}^{2}(v,u,x)} P_{\zeta}(kv)P_{\zeta}(ku)
		\label{eq:phuv}
	\end{equation} 
	where $x\equiv k\eta$.
	In the late time limit $x\rightarrow \infty$, 
	the function $\overline{\mathrm{I}^{2}(v,u,x)}$ is obtained as \cite{Kohri:2018awv}
	\begin{align}
		&\overline{\mathrm{I}^{2}(v,u,x\rightarrow \infty)} =  \frac{9}{2x^{2}} \left(\frac{u^{2}+v^{2}-3}{4u^{3}v^{3}} \right)^{2} \times \nonumber \\ &\bigg[\left(-3uv+ (u^{2}+v^{2}-3)\log  \left| \frac{3-(u+v)^{2}}{3-(u-v)^{2}}\right|\right)^{2}   +  \pi^{2}(u^{2}+v^{2}-3)^{2}\theta(u+v-\sqrt{3}) 
		\bigg].
		\label{IRD}
	\end{align}
	Having obtained the expression for gravitational wave energy density in Eq. (\ref{eq:omegagw}), the observational relevant quantity, the present energy spectrum of secondary gravitational waves, $\Omega_{\mathrm{GW},0}(k)$ is estimated as \cite{Bastero-Gil:2021fac}
	\begin{equation}
		\Omega_{\mathrm{GW},0}(k)=0.39\left(\frac{g_{\star}}{106.75} \right)^{-\frac{1}{3}} \Omega_{r,0}~ \Omega_{\mathrm{GW}}(\eta_{c},k)
		\label{eq:omegagw0}
	\end{equation}
	where $g_{\star} $ is the effective number of relativistic degrees of freedom in the radiation-dominated era and $\Omega_{r,0} $ is the present radiation energy density. Here, $\eta_{c}$ represents the conformal time at an epoch when a perturbation is inside the horizon after re-entry during radiation dominated era. To estimate $\Omega_{\mathrm{GW},0}(k)$ as of frequency, we replace $k$ with frequency, $f$ from the relation
	\begin{equation}
		f=\frac{k}{2\pi}=1.5\times 10^{-15}\left( \frac{k}{1~ \rm{Mpc}^{-1}}\right)  \rm{Hz}.
		\label{eq:frkrel}
	\end{equation}

	\section{Results}
	\label{results}
	
	After describing the basic formalism of PBH formation and the generation of 
	secondary gravitational waves,  we now discuss the results obtained for our warm Higgs-G inflation model. As discussed before, we will work with the parameter space for $Q_G$ value at the pivot scale (denoted by $Q_P$) (and the corresponding $C_T$, $M$ values) consistent with the CMB observations.

	\subsection{Primordial curvature power spectrum}
	\label{PPS}
	
	Using the parameterization of our warm Higgs-G model, as carried out in Section \ref{WHGI}, we evolve $Q_G$ as a function of the comoving scale $k$ or number of efolds $N_e$, and obtain the evolution of the primordial power spectrum. Here we show the results of our calculations.
	
\begin{figure}[]
		\subfigure[]{
			\label{fig:2a}
			\includegraphics[scale=0.6]{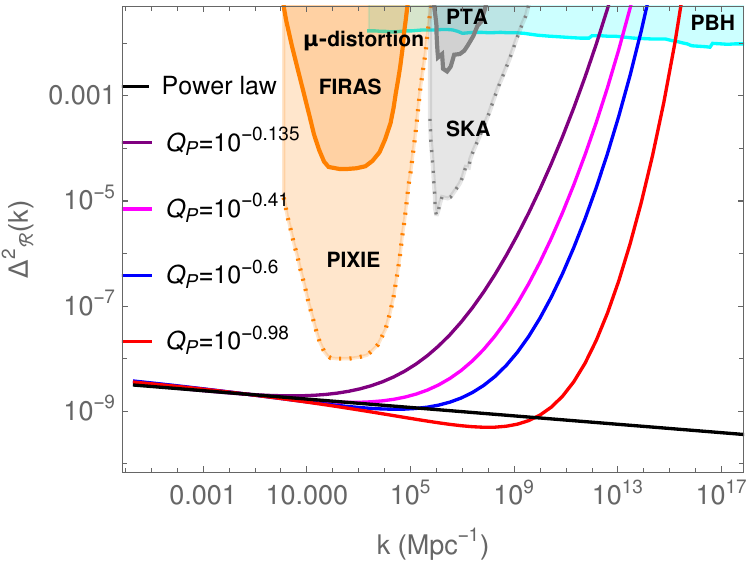}}
		\hspace{0.1cm} 
		\subfigure[]{
			\label{fig:2b}
			\includegraphics[scale=0.6]{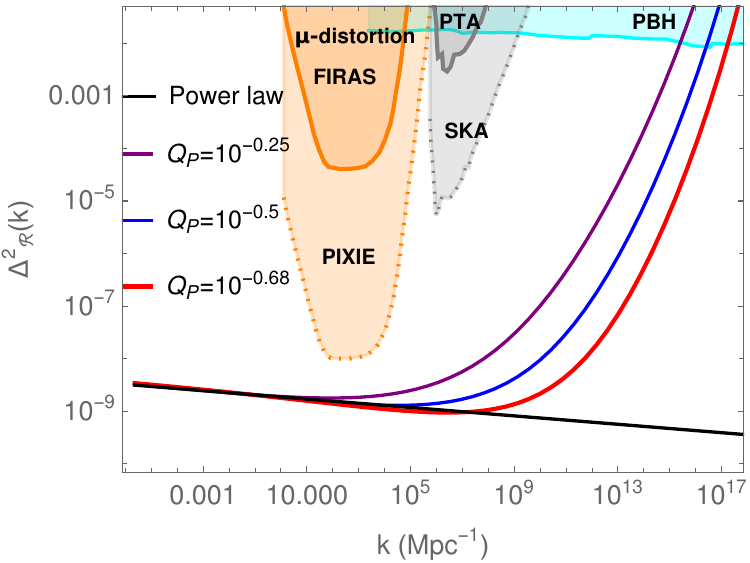}} 
		\caption{The plots of primordial curvature power spectrum as a function of comoving scale $k$ for the different values of $Q_G$ allowed by CMB for Model I ($p=0, q=1$). Fig. \ref{fig:2a} and Fig. \ref{fig:2b} correspond to $50$  and $60$ efolds of inflation, respectively. The black solid line represents the usually considered power-law primordial power spectrum. The observational constraints from $\mu-$distortion, PTA and SKA are taken from Ref. \cite{Chluba:2019kpb}. 
			The cyan-shaded region represents the excluded region from PBH overproduction.
		}
		\label{p0q0}
	\end{figure}

\begin{figure}[]
		\subfigure[]{%
			\label{fig:31a}%
			\includegraphics[scale=0.6]{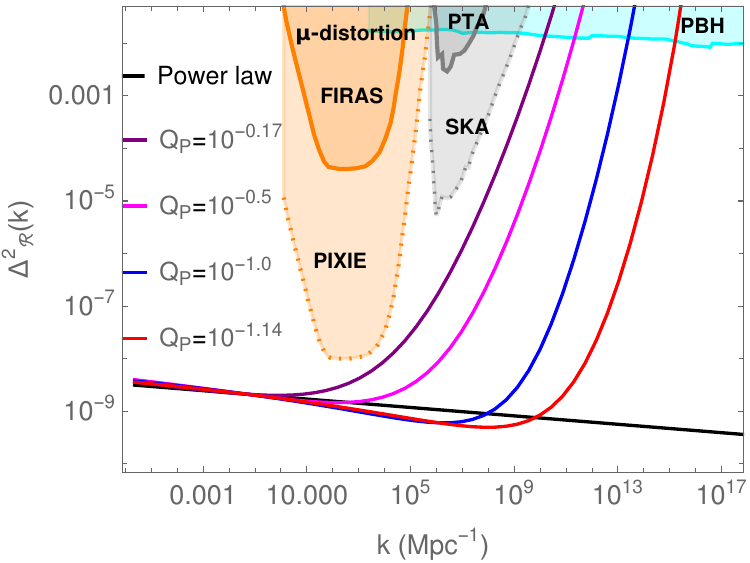}} 
		\hspace{0.1cm}
		\subfigure[]{%
			\label{fig:31b}%
			\includegraphics[scale=0.6]{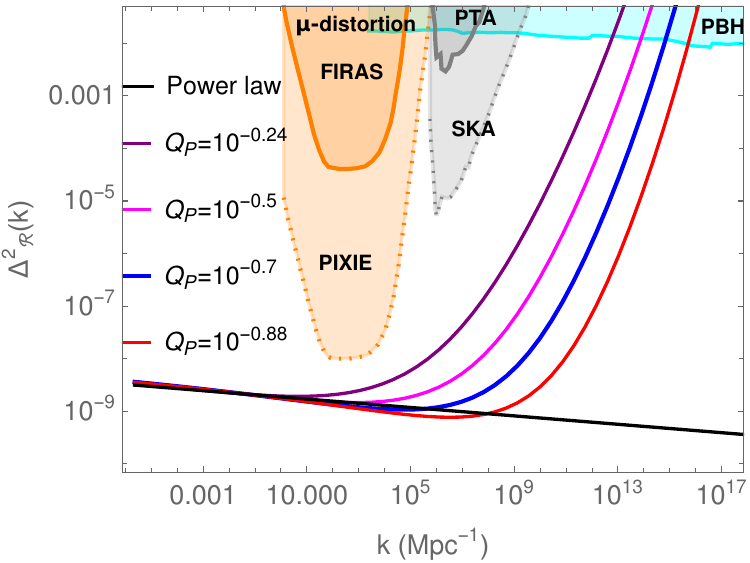}} 	
		\caption{The plots of primordial curvature power spectrum as a function of $k$ for the different values of $Q_G$ allowed by CMB for Model II ($p=1, q=1$). Fig. \ref{fig:31a} and Fig. \ref{fig:31b} correspond to $50$  and $60$ efolds of inflation, respectively. The black solid line represents the usually considered power-law primordial power spectrum. The observational constraints from $\mu-$distortion, PTA and SKA are taken from Ref. \cite{Chluba:2019kpb}.  The cyan-shaded region represents the excluded region from PBH overproduction. 
		}
		\label{p1q1}
	\end{figure}

	We plot the primordial curvature power spectrum as a function of  $k$ in Fig. \ref{p0q0} and Fig. \ref{p1q1} for Model I ($p=0, q=1$) and Model II ($p=1, q=1$), respectively. In these figures, we show the results for the allowed range of $Q_G$ values (at the pivot scale) for both $50$ and $60$ efolds of inflation.  It can be seen that small variations in the initial $Q_G$ values can evolve differently and lead to tremendous growth at different comoving scales.
	Also, we show the observational constraints on the power spectrum obtained from $\mu-$distortion, PTA and SKA, taken from Ref. \cite{Chluba:2019kpb}.
	The cyan-shaded region represents the excluded region from PBH overproduction.
	The black solid line in these plots represents the standard power-law form of primordial power spectrum, $\Delta^2_\mathcal{R} (k)=A_s\left(\frac{k}{k_p}\right)^{n_s-1}$, which is red-tilted ($n_s<1$) at all the scales, thus cannot form PBHs.
	Whereas in both Fig. \ref{p0q0} and Fig. \ref{p1q1}, we can see that the spectrum for warm inflation is red-tilted at the CMB scales, and it turns blue-tilted ($n_s>1$) at the small scales with a huge enhancement in the amplitude. As we increase the value of $Q_P$, we see that the power spectrum rises steeper and attains the value $\sim 10^{-2}$ at a comparatively smaller $k$ value.  As discussed before, PBHs of an observationally significant abundance are formed when the amplitude of the primordial power spectrum is of the order of $\Delta^2_\mathcal{R} (k)\sim \mathcal{O} (10^{-2})$. 
	
	We see from Fig. \ref{fig:2a}, for the case of $50$ efolds of Model I ($p=0,q=1$), this is achieved for $1.37\times 10^{12}<k<1.5\times 10^{15} $ Mpc$^{-1}$, while for $60$ efolds, as shown in Fig. \ref{fig:2b}, this condition is achieved for $2.2\times 10^{15}<k<1.78\times 10^{17} $ Mpc$^{-1}$. 
	Similarly, from Fig. \ref{fig:31a} for Model II ($p=1,q=1$), we find that for 50 efolds, the amplitude of primordial power spectrum is of order $10^{-2}$ 
	for scales\footnote{For our further calculations, we take these cutoff values of comoving wavenumbers at which primordial power spectrum $\sim \mathcal{O}(10^{-2})$. Similar approach was also taken in Ref. \cite{Kohri:2018qtx}.} $1.14\times 10^{10}<k<1.46\times 10^{15} $ Mpc$^{-1}$.
	For $60$ efolds, we have a sufficient amplitude of power on the scales $4.2\times 10^{12}<k<4.8\times 10^{15} $ Mpc$^{-1}$, as shown in Fig. \ref{fig:31b}.
	
	We would like to point out that in a canonical setup, the $\lambda\phi^4$ warm inflation model with linear dissipation coefficient $\Upsilon\propto T$ does not lead to a sufficient enhancement of the primordial power, while a cubic dissipation $\Upsilon\propto T^3$ could lead to PBH generation \cite{Arya:2019wck}.
	However, in the non-canonical model, even the linear dissipation coefficient is sufficient to cause enough enhancement in the power spectrum and leads to PBH formation. As discussed previously, the spectral index is highly blue-tilted for the small scales, as a result of which the power spectrum keeps growing, which might raise a concern. However, we propose that this issue can be resolved if we have a model, for example, in which the dissipation coefficient switches its behaviour from $\Upsilon\propto T$ (for $30-40$ efolds) to $\Upsilon\propto \phi^2/T$ ($10-20$ efolds) as inflation proceeds.  The slow-roll evolution of the inflaton field will be different in the two phases and  the full background evolution can be studied numerically, incorporating both forms of the dissipation coefficient.
Note that the range of $Q_P$ allowed from the spectral index observations, as shown in Section \ref{Qprange}, will be different in this proposed model \footnote{We find the valid parameter range as $0.016<Q_G<0.8$ (for $p=0, q=1$ model) and  $0.02<Q_G<0.6$ (for $p=1, q=1$ model), considering $30$ efolds in the linear dissipative regime and $20$ efolds in the $\phi^2/T$ dissipative regime.} and thus, that particular range 
should be studied for the PBH formation.
The dissipation coefficient  $\Upsilon\propto \phi^2/T$  is also considered in Refs. \cite{Hosoya:1983ke,Hall:2003zp}. The crucial point about this form of $\Upsilon$ is that the $G(Q) (c<0)$ factor leads to suppression in the primordial power spectrum \footnote{Note that $\Upsilon\propto 1/T$ could also lead to suppression in the power spectrum, but in the regime $Q_G\gg1$, inflation does not come to an end when $\Upsilon\propto1/T$.}. Hence, if such a feature is introduced after the peak enhancement, this can resolve the issue of the validity of the linear analysis.
We have not carried out a rigorous calculation of the primordial power spectrum for our proposed model. It will require solving the coupled differential equations for inflaton-radiation system numerically, as done in Refs. \cite{Graham:2009bf, Bastero-Gil:2021fac,Ballesteros:2022hjk}, and will be presented in a future work.
However, the crucial finding of our present study is the enhancement in the power spectrum due to a non-canonical term in inflaton evolution, leading to the formation of PBHs as the entire dark matter. 
	
	\subsection{Initial mass fraction of PBHs}	
\begin{figure}[]
		\subfigure[]{%
			\label{fig:3a}%
			\includegraphics[scale=0.64]{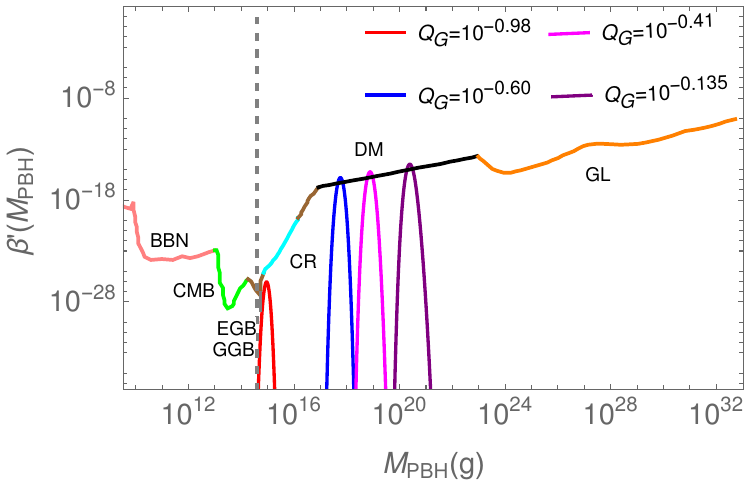}}%
		\hspace{0.2cm}
		\subfigure[]{%
			\label{fig:3b}%
			\includegraphics[scale=0.64]{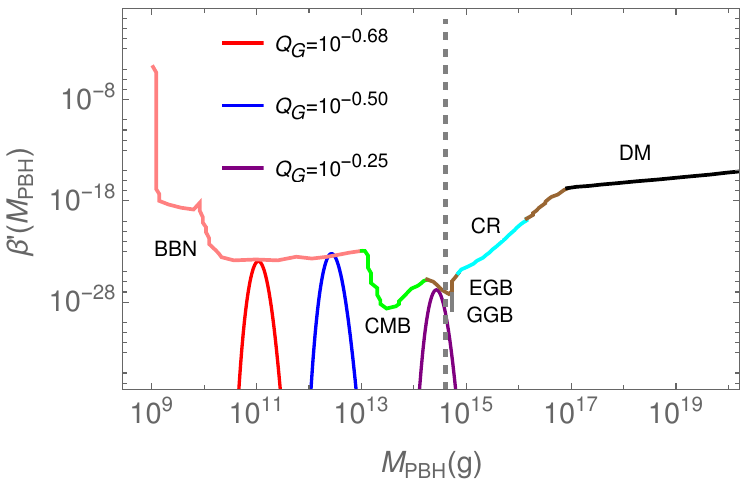}} 
		\caption{Plots of $\beta'(M_{PBH})$ versus $M_{PBH}$ using allowed parameter space of $Q_G$ for \ref{fig:3a}: 50 efolds of inflation, and \ref{fig:3b}: 60 efolds of inflation  in Model I ($p=0, q=1$). The different observational constraints are taken from Ref. \cite{ Carr:2020gox}. Here the acronyms stand for: BBN, CMB, extragalactic $\gamma$-ray background (EGB), galactic $\gamma$-ray background (GGB), cosmic ray (CR), DM,  gravitational lensing (GL). The constraints here are for the monochromatic mass function of PBHs.}
		\label{beta-p0q1}
	\end{figure}

\begin{figure}[]
		\subfigure[]{%
			\label{fig:4a}%
			\includegraphics[scale=0.64]{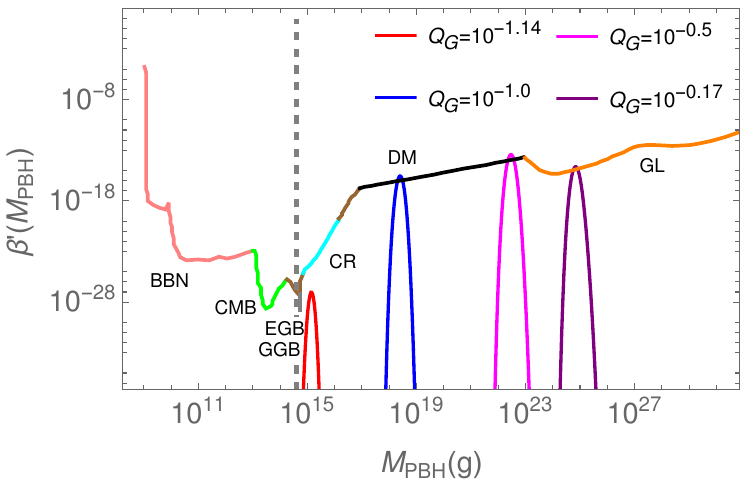}}%
		\hspace{0.2cm}
		\subfigure[]{%
			\label{fig:4b}%
			\includegraphics[scale=0.64]{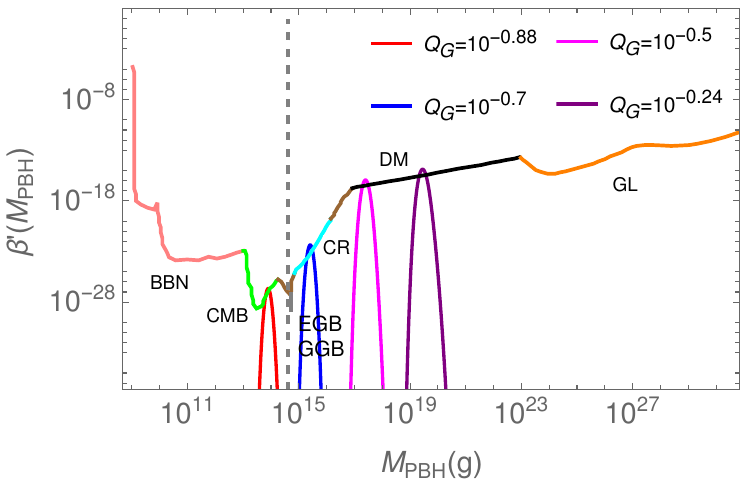}} 
		\caption{Plots of $\beta'(M_{PBH})$ versus $M_{PBH}$ using allowed parameter space of $Q_G$ for \ref{fig:4a}: 50 efolds of inflation, and \ref{fig:4b}: 60 efolds of inflation in Model II ($p=1, q=1$). The different observational constraints are taken from Ref. \cite{Carr:2020gox}. The acronyms are the same as in the previous figure. }
		\label{beta-p1q1}
	\end{figure}

	Till now, we have identified the range of comoving scales for our warm inflation models, which will be of interest to us. When these modes reenter the horizon during the radiation-dominated era, they collapse into PBHs.
	We now show the results for the mass and abundance of the generated PBHs from our models, using the formalism discussed in Section \ref{PBH}.  
	We plot the initial mass fraction, $\beta'(M_{PBH})$ (defined equal to $\gamma^{1/2} \left(\frac{g_*}{106.75}\right)^{-1/4} \left(\frac{h}{0.67}\right)^{-2} \beta(M_{PBH})$) versus $M_{PBH}$ for our WI models in Fig. \ref{beta-p0q1} and Fig. \ref{beta-p1q1} corresponding to Model I ($p=0, q=1$) and Model II ($p=0, q=1$), respectively. We infer from these plots that the larger the $Q_G$ value, the more massive the PBHs are. This is because for larger $Q_G$, the growth of primordial power is steeper and the large amplitude is attained at a smaller $k$ value. As $M_{PBH}$ is inversely proportional to $k$, this implies a more massive PBH generation. For a review on the observational constraints on $\beta'(M_{PBH})$, see Ref. \cite{ Carr:2020gox}.

	We can see from Fig. \ref{fig:3a} that for 50 efolds of inflation of Model I, PBHs over a mass range $9.3\times10^{14}$ g $<M_{PBH}<5.6\times10^{20}$ g can be generated, corresponding to the scales quoted in the previous subsection. Some part of this range lies in the interesting asteroid mass window, thus leading to a  possibility of explaining the full dark matter abundance.
	For $60$ efolds of inflation of Model I, the mass of generated PBHs varies from $9.5\times10^{10}$ g $<M_{PBH}<4\times10^{14}$ g, as shown in Fig. \ref{fig:3b}. These PBHs would have evaporated into Hawking radiation by today and are thus constrained through the PBH evaporation bounds.
	
	Similarly, in Fig. \ref{beta-p1q1}, we plot $\beta'(M_{PBH})$ versus $M_{PBH}$ using  Model II ($p=1, q=1$) for both $50$ and $60$ efolds of inflation. We find that for 50 efolds, shown in Fig. \ref{fig:4a}, PBHs over a mass range $1.2\times10^{15}$ g $<M_{PBH}<2\times10^{25}$ g can be generated, while for $60$ efolds of inflation, mass of the produced PBHs lies between $1\times10^{14}$ g $<M_{PBH}<9.8\times10^{19}$ g, as can be seen in Fig. \ref{fig:4b}. Therefore, we emphasize that in both scenarios, our model predicts a possibility that PBHs might constitute the total dark matter abundance.

	\subsection{PBHs as dark matter}
\begin{figure}[]
		\centering
		\includegraphics[scale=0.7]{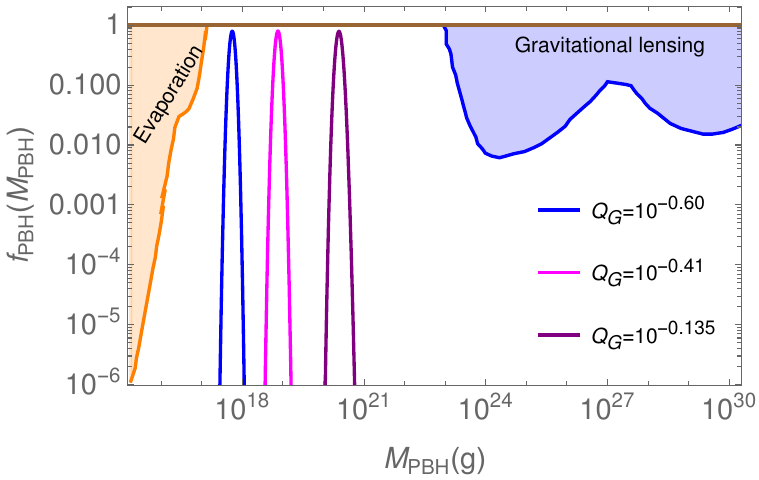}
		\caption{Fraction of DM in PBHs, $f_{PBH}$  vs.  $M_{PBH}$ using the allowed parameter space of $Q_G$ for 50 efolds of inflation in Model I ($p=0, q=1$). The evaporation and gravitational lensing constraints are taken from Ref. \cite{Carr:2020xqk}. The maximum fraction in PBHs is bounded from above as $f_{PBH}\leq 1$.} 
		\label{dm-p0q1}
	\end{figure}

\begin{figure}[]
		\subfigure[]{%
			\label{fig:a}%
			\includegraphics[scale=0.62]{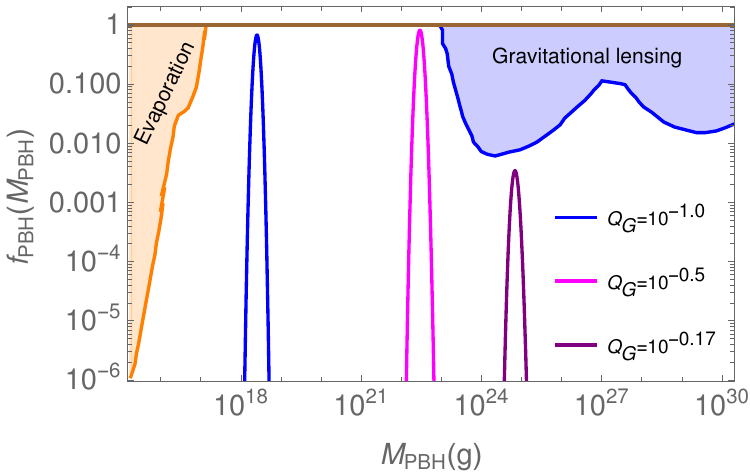}}
		\hspace{0.2cm}%
		\subfigure[]{%
			\label{fig:b}%
			\includegraphics[scale=0.62]{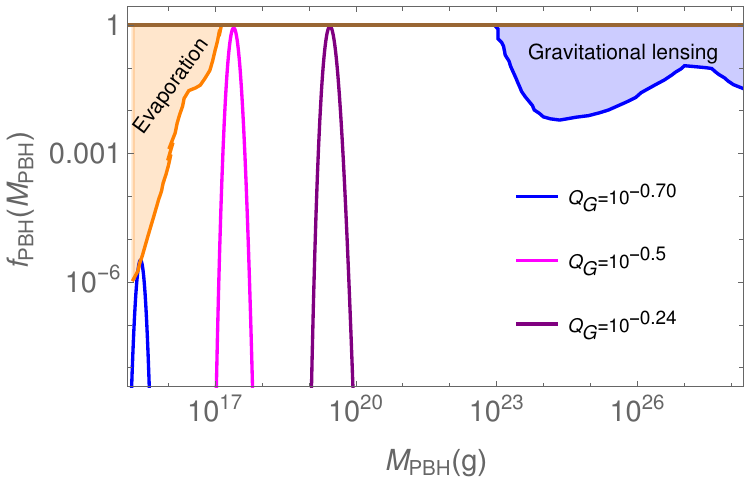}} 
		\caption{Fraction of dark matter in PBHs, $f_{PBH}$  versus $M_{PBH}$ using the allowed parameter space of $Q_G$ for \ref{fig:a}: 50 efolds of inflation,  and \ref{fig:b}: 60 efolds of inflation  in Model II ($p=1, q=1$).  The evaporation and gravitational lensing constraints are taken from Ref. \cite{Carr:2020xqk}.  }
		\label{dm-p1q1}
	\end{figure}

	PBHs can contribute significantly or total fraction of the present day DM energy density in certain mass ranges. Here we explore this possibility for our warm inflation models.
	For this, we calculate and plot the fraction of DM in PBHs, $f_{PBH}$ as a function of PBH mass in Fig. \ref{dm-p0q1} and Fig. \ref{dm-p1q1} using the formalism in Section \ref{PBH}. In these plots, we focus on the PBHs mass range greater than $\sim 10^{15}$ g, because the smaller mass PBHs would have evaporated through Hawking radiation and, therefore, will not contribute much to the DM.  
	The PBHs can explain the full DM when $f_{PBH}=1$. 
	From Fig. (\ref{dm-p0q1}), we see that in Model I ($p=0, q=1$) with $50$ efolds of inflation, there is a production of the asteroid mass range PBHs, which can explain the full DM abundance. 
	We further point out that this model also predicts smaller mass PBHs, with low DM fractional abundance.
	
	Similarly, Fig. \ref{fig:a} indicates that Model II ($p=1, q=1$) with $50$ efolds also generates asteroid mass range PBHs that can constitute the full DM abundance. Along with that, this model also leads to comparably higher and lower mass PBHs, and is constrained through microlensing or evaporation observations, respectively. Further, from Fig. \ref{fig:b}, we infer that for $60$ efolds, this model also leads to a possibility of the production of PBH DM in the asteroid mass range. This model also predicts smaller mass PBHs which do not contribute significantly to the DM, but are consistent with the evaporation bounds.
	

	\subsection{Spectrum of scalar induced gravitational waves }
\begin{figure}[t]
		\subfigure[]{%
			\label{fig:7a}%
			\includegraphics[scale=0.62]{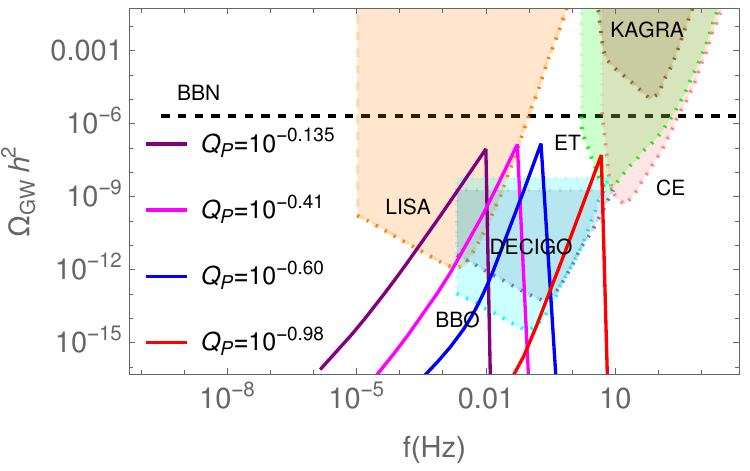}}%
		\hspace{0.2cm}
		\subfigure[]{%
			\label{fig:7b}%
			\includegraphics[scale=0.62]{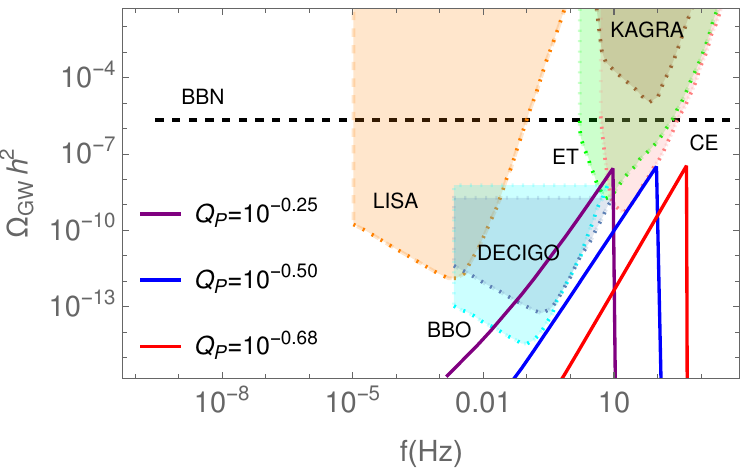}} 
		\caption{Plot of the gravitational wave energy density, $\Omega_{GW} h^{2}$ induced by the scalar perturbations as a function of frequency for the allowed parameter space of $Q_G$ for \ref{fig:7a}: 50 efolds of inflation, and \ref{fig:7b}: 60 efolds of inflation, for Model I ($p=0, q=1$). The sensitivity plots for future GW detectors are taken from Ref. \cite{Kohri:2018awv}. The constraints here are for the monochromatic mass function of PBHs. }
		\label{gw-p0q1}
	\end{figure}

\begin{figure}[]
		\subfigure[]{%
			\label{fig:8a}%
			\includegraphics[scale=0.62]{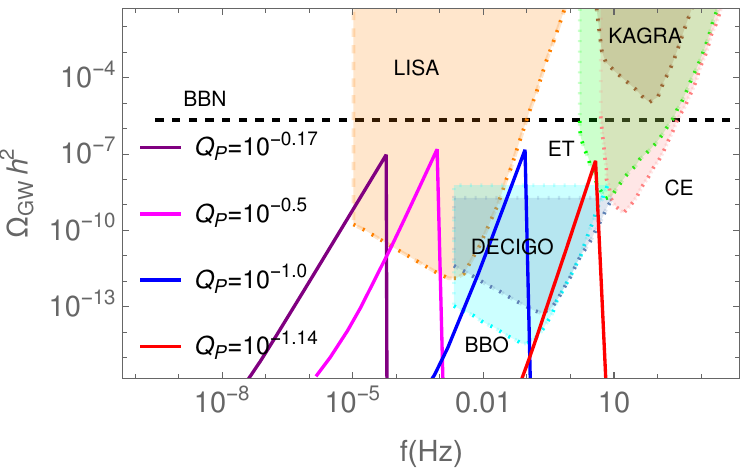}}%
		\hspace{0.2cm}
		\subfigure[]{%
			\label{fig:8b}%
			\includegraphics[scale=0.62]{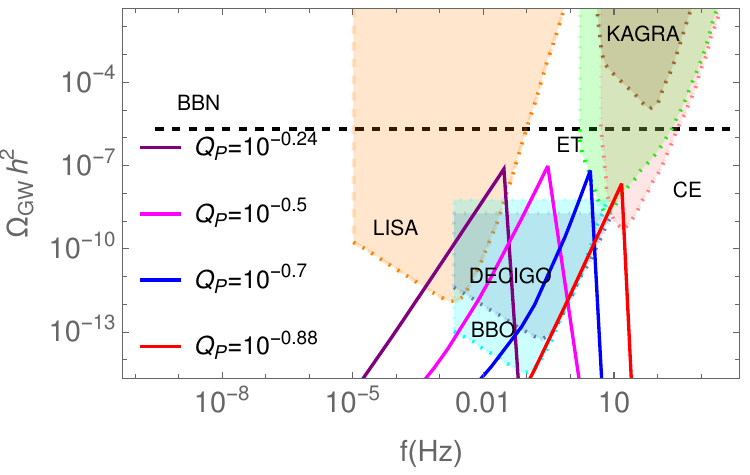}} 
		\caption{Plot of the gravitational wave energy density, $\Omega_{GW} h^{2}$ induced by the scalar perturbations as a function of frequency for the allowed parameter space of $Q_G$ for \ref{fig:8a}: 50 efolds of inflation, and \ref{fig:8b}: 60 efolds of inflation, for Model II ($p=1, q=1$). The sensitivity plots for future GW detectors are taken from Ref. \cite{Kohri:2018awv}. }
		\label{gw-p1q1}
	\end{figure}

	The large density fluctuations needed for PBH formation inevitably lead to secondary gravitational wave generation, as discussed before. From Eqs. (\ref{Mpbh}) and (\ref{eq:frkrel}), we see that the peak GW frequency inversely depends on the PBHs mass. The larger the mass of PBH, the smaller the associated peak GW frequency. To explore the spectrum of scalar-induced gravitational waves, we plot $\Omega_{GW} h^{2}$ as a function of frequency 
	in Fig. \ref{gw-p0q1} and Fig. \ref{gw-p1q1} 
	corresponding to the  Model I ($p=0, q=1$) and Model II ($p=1, q=1$), respectively. The sensitivity plots for future GW detectors are taken from Ref. \cite{Kohri:2018awv}.
	
	From Fig. \ref{fig:7a}, we see that Model I ($p=0, q=1$) with 50 efolds of inflation produces $9.3\times10^{14}$ g $<M_{PBH}<5.6\times10^{20}$ g mass PBHs, and generates a GW of peak frequency lying between $\sim (10^{-2}$ $-10$) Hz. These can be detected in future GW detectors such as LISA, BBO, DECIGO, ET, and CE. Further, the same model
	with 60 efolds of inflation produces comparably lower PBH mass range, $9.5\times10^{10}$ g $<M_{PBH}<4\times10^{14}$ g, and generates comparably higher peak frequency of GW lying between $\sim (10-500$) Hz, as can be seen in Fig. \ref{fig:7b}. These can be detected in the future GW detector sensitive to high frequencies such as ET and CE.

	Similarly, Fig. \ref{fig:8a} suggests that Model II ($p=1, q=1$) with 50 efolds of inflation produces $1.2\times10^{15}$ g $<M_{PBH}<2\times10^{25}$ g mass PBHs and generates GW of peak frequencies lying between $\sim (10^{-4}-10$) Hz. These can be detected in GW observations such as LISA, BBO, DECIGO, and ET. Further, the same model with 60 efolds of inflation produces PBH mass range, $1\times10^{14}$ g $<M_{PBH}<9.8\times10^{19}$ g, which generates a peak frequency of GW lying between $\sim (0.1 -30$) Hz, as shown in Fig. \ref{fig:8b}. These can also be detected in future GW detectors such as LISA, BBO, DECIGO, ET, and CE, and thus used to test these models of inflation.

	\section{Summary and Discussion}
	\label{summary}
	
	The inflationary paradigm of the early Universe has been extremely successful in explaining various cosmological observations, however, the underlying particle physics model to describe this accelerating phase is not clearly known. The present demand of the model building is not only to construct viable models with a physical motivation and interesting phenomenology but also successfully embed them in a UV complete theory.
	The framework of warm inflation is a general and well-motivated description of inflation wherein the dissipative processes in a coupled inflaton-radiation system drive the evolution of the early Universe.
	The inflaton dissipates its energy into radiation fields, as a result of which there is a non-zero temperature in the Universe throughout the inflationary phase. The inflaton background evolution, as well as its perturbations are modified due to the presence of the thermal bath.
	Thus, warm inflation leads to unique signatures on the large and small scale observations, and hence important to study.
	
	In this paper, we have studied the scenario of warm Higgs-G inflation, wherein the Standard Model Higgs boson plays the role of the inflaton, and has a Galileon-like non-linear kinetic term, which contributes as an additional frictional term in its evolution. For the quartic Higgs potential and a dissipation coefficient linear in temperature, we have studied two different cases of warm Higgs-G inflation models, with $G(\phi, X) \propto \phi^{2p+1}X^q$ and parameter sets ($p=0,q=1$) and ($p=1,q=1$). For a wide range of other parameters, we found that these models are compatible with the CMB Planck observations at large scales. Moreover, contrary to the usual cold inflation, we have shown that these scenarios are consistent with the swampland and the TCC conjectures, implying that they lie in the viable landscape of UV complete theories. 
	In these scenarios, the non-linear kinetic term plays an important role, and we found that when it dominates, the background dynamics of the inflaton and the evolution of small scale perturbations are modified. This leads to a blue-tilted spectrum with an enhanced amplitude at small scales, inducing the generation of PBHs and their associated observational imprints, such as the induced GWs.
	
	In our analysis, we found that PBHs over a wide mass range are generated in these models, in particular, in the asteroid mass range $(10^{17}-10^{23})$ g, which can explain the total dark matter abundance at present. 
	This particular mass range is interesting as it remains unconstrained from observations, and therefore, PBHs produced in any scenario can be the entire dark matter only in this mass range, while in other mass ranges, they can, at most, be a fraction.  
	We have further calculated the secondary scalar induced GWs sourced by these small scale overdense fluctuations in our set-up and found that the induced GW spectrum can be detected in future GW detectors, such as LISA, BBO, DECIGO, ET and CE. In particular, for some GWs spectra, there also exists an interesting possibility of their simultaneous detection with many GW observatories. We conclude that warm inflationary models have interesting prospective signatures, and the forthcoming cosmological probes would be very useful for testing these models.

	Furthermore, it is well known that primordial non-Gaussianities strongly affect the abundance of PBHs as they form from the tail of the density distribution \cite{Cai:2018dig,Unal:2018yaa,Ragavendra:2021qdu}. Studies show that in warm inflation, the amplitude and shapes of non-Gaussianities are very different in the strong and weak dissipative regimes.  
	It will therefore be interesting to include and understand the effects of such non-Gaussianities in the warm inflationary models for the calculation of PBH abundance as well as for the induced GW background.  We will investigate these interesting aspects in future works.

	\section*{Acknowledgements}
	RA acknowledges the support from the National Post-Doctoral
	Fellowship by the Science and Engineering Research Board (SERB), Department of Science and Technology (DST), Government of India (GOI)(PDF/2021/004792). The work of AKM is supported through Ramanujan Fellowship
	(PI: Dr. Diptimoy Ghosh) offered by the DST, GOI (SB/S2/RJN-088/2018).
	RKJ acknowledges financial support from the new faculty seed start-up 
	grant of the Indian Institute of Science, Bengaluru, India,  SERB, DST, GOI,
	through the Core Research Grant~CRG/2018/002200, the MATRICS grant~MTR/2022/000821 and the Infosys Foundation, Bengaluru, India through the Infosys Young Investigator Award. We also acknowledge the anonymous referee for providing critical and insightful comments that led to an improvement in the presentation of the paper.
	
	\bibliographystyle{utphys}
	\bibliography{ref}
\end{document}